\documentclass[a4paper,fleqn]{cas-dc}

\PassOptionsToPackage{hyphens}{url} 
\usepackage{hyperref}
\usepackage{amsmath,amssymb,amsfonts}

\usepackage{textcomp}
\usepackage[utf8]{inputenc}
\usepackage{xcolor}
\usepackage{lscape}
\usepackage{pdflscape}
\usepackage{graphicx}
\usepackage{booktabs}
\usepackage{dirtytalk}
\usepackage{longtable}
\usepackage{wasysym}
\usepackage{multirow}
\usepackage[inline]{enumitem}
\usepackage[utf8]{inputenc}
\usepackage{csquotes}
\usepackage{tikz}
\usetikzlibrary{turtle}
\usepackage{acro}
\usepackage[english]{babel}
\usepackage{tabularx}
\usepackage{longtable}
\usepackage{colortbl}
\usepackage{array}
\usepackage{makecell}
\usepackage[normalem]{ulem}
\useunder{\uline}{\ul}{}
\usepackage{colortbl}

\usepackage{algorithm}
\usepackage{algpseudocode}
\usepackage{xcolor}

\usepackage{pifont}

\usepackage{placeins}

\usepackage[numbers]{natbib}

\usepackage{rotating}

\usepackage{xurl}

\usepackage{etoolbox}

\usepackage{needspace}

\usepackage[acronym]{glossaries-extra}
\makeglossaries

\def\tsc#1{\csdef{#1}{\textsc{\lowercase{#1}}\xspace}}
\tsc{WGM}
\tsc{QE}
\tsc{EP}
\tsc{PMS}
\tsc{BEC}
\tsc{DE}

\NewDocumentCommand{\rot}{O{45} O{1em} m}{\makebox[#2][l]{\rotatebox{#1}{#3}}}

\newtheorem{definition}{Definition}[section]

\newcommand{\oldstuff}[1]{}
\newcommand{\old}[1]{}

\newcommand*\emptycirc[1][1ex]{\tikz\draw (0,0) circle (#1);} 
\newcommand*\halfcirc[1][1ex]{%
  \begin{tikzpicture}
  \draw[fill] (0,0)-- (90:#1) arc (90:270:#1) -- cycle ;
  \draw (0,0) circle (#1);
  \end{tikzpicture}}
\newcommand*\fullcirc[1][1ex]{\tikz\fill (0,0) circle (#1);}

\newcommand{\smallfullcirc}{\raisebox{-0.1em}{\scalebox{0.75}{\fullcirc}}}
\newcommand{\smallemptycirc}{\raisebox{-0.1em}{\scalebox{0.75}{\emptycirc}}}
\newcommand{\smallhalfcirc}{\raisebox{-0.1em}{\scalebox{0.75}{\halfcirc}}}

\definecolor{r0}{rgb}{1, 1.00, 1.00}
\definecolor{r1}{rgb}{1, 0.90, 0.90}
\definecolor{r2}{rgb}{1, 0.80, 0.80}
\definecolor{r3}{rgb}{1, 0.70, 0.70}
\definecolor{r4}{rgb}{1, 0.60, 0.60}
\definecolor{r5}{rgb}{1, 0.50, 0.50}
\definecolor{r6}{rgb}{1, 0.40, 0.40}
\definecolor{r7}{rgb}{1, 0.30, 0.30}
\definecolor{r8}{rgb}{1, 0.20, 0.20}
\definecolor{r9}{rgb}{1, 0.10, 0.10}
\definecolor{r10}{rgb}{1, 0.00, 0.00}
\definecolor{g0}{rgb}{1.00, 1, 1.00}
\definecolor{g1}{rgb}{0.90, 1, 0.90}
\definecolor{g2}{rgb}{0.80, 1, 0.80}
\definecolor{g3}{rgb}{0.70, 1, 0.70}
\definecolor{g4}{rgb}{0.60, 1, 0.60}
\definecolor{g5}{rgb}{0.50, 1, 0.50}
\definecolor{g6}{rgb}{0.40, 1, 0.40}
\definecolor{g7}{rgb}{0.30, 1, 0.30}
\definecolor{g8}{rgb}{0.20, 1, 0.20}
\definecolor{g9}{rgb}{0.10, 1, 0.10}
\definecolor{g10}{rgb}{0.00, 1, 0.00}

\definecolor{o0}{rgb}{1.00, 0.90, 0.70}  
\definecolor{o1}{rgb}{1.00, 0.85, 0.55}
\definecolor{o2}{rgb}{1.00, 0.75, 0.40}
\definecolor{o3}{rgb}{1.00, 0.65, 0.25}
\definecolor{o4}{rgb}{1.00, 0.55, 0.10}
\definecolor{o5}{rgb}{1.00, 0.50, 0.00}  
\definecolor{o6}{rgb}{0.90, 0.45, 0.00}
\definecolor{o7}{rgb}{0.80, 0.40, 0.00}
\definecolor{o8}{rgb}{0.70, 0.35, 0.00}
\definecolor{o9}{rgb}{0.60, 0.30, 0.00}
\definecolor{o10}{rgb}{0.50, 0.25, 0.00} 

\newcommand{\orange}[1]{\textcolor{orange}{#1}}

\newcommand{\teal}[1]{\textcolor{teal}{#1}}
\newcommand{\olive}[1]{\textcolor{olive}{#1}}

\makeatletter
\let\disable\@secondoftwo
\addto\extrasenglish{%
  \renewcommand{\sectionautorefname}{\S\@gobble}%
}	
\addto\extrasenglish{%
  \renewcommand{\subsectionautorefname}{\S\@gobble}%
}
\addto\extrasenglish{%
  \renewcommand{\subsubsectionautorefname}{\S\@gobble}%
}
\addto\extrasenglish{%
  \renewcommand{\paragraphautorefname}{\S\@gobble}%
}
\makeatother

\AtBeginDocument{%
  \providecommand\BibTeX{{%
    \normalfont B\kern-0.5em{\scshape i\kern-0.25em b}\kern-0.8em\TeX}}}

\DeclareAcronym{aum}{
  short = AUM,
  long  = assets under management
}

\DeclareAcronym{hmac}{
  short = HMAC,
  long  = hash-based message authentication code
}
\DeclareAcronym{aes}{
  short = AES,
  long  = Advanced Encryption Standard
}
\DeclareAcronym{sha}{
  short = SHA,
  long  = Secure Hash Algorithm
}

\DeclareAcronym{gcm}{
  short = GCM,
  long  = Galois/Counter Mode
}

\DeclareAcronym{slip}{
  short = SLIP,
  long  = SatoshiLabs Improvement Proposal
}

\DeclareAcronym{derec}{
  short = DeRec,
  long  = Decentralised Recovery
}

\DeclareAcronym{bip}{
  short = BIP,
  long  = Bitcoin Improvement Proposal
}

\DeclareAcronym{eip}{
  short = EIP,
  long  = Ethereum Improvement Proposal
}

\DeclareAcronym{erc}{
  short = ERC,
  long  = Ethereum Request for Comments
}

\DeclareAcronym{evm}{
  short = EVM,
  long  = Ethereum Virtual Machine
}

\DeclareAcronym{cve}{
  short = CVE,
  long  = Common Vulnerabilities and Exposures
}

\DeclareAcronym{cvss}{
  short = CVSS,
  long  = Common Vulnerability Scoring System
}

\DeclareAcronym{ens}{
  short = ENS,
  long  = Ethereum Name Service
}

\DeclareAcronym{bgp}{
  short = BGP,
  long  = Border Gateway Protocol
}

\DeclareAcronym{arp}{
  short = ARP,
  long  = Address Resolution Protocol
}

\DeclareAcronym{icmp}{
  short = ICMP,
  long  = Internet Control Message Protocol
}

\DeclareAcronym{tcp}{
  short = TCP,
  long  = Transmission Control Protocol
}

\DeclareAcronym{ip}{
  short = IP,
  long  = Internet Protocol
}

\DeclareAcronym{dns}{
  short = DNS,
  long  = Domain Name System
}

\DeclareAcronym{nfc}{
  short = NFC,
  long  = Near Field Communication
}

\DeclareAcronym{ecc}{
  short = ECC,
  long  = Elliptic Curve Cryptography
}

\DeclareAcronym{ecdsa}{
  short = ECDSA,
  long  = Elliptic Curve Digital Signature Algorithm
}

\DeclareAcronym{eddsa}{
  short = EdDSA,
  long  = Edwards-curve Digital Signature Algorithm
}

\DeclareAcronym{dsa}{
  short = DSA,
  long  = Digital Signature Algorithm
}

\DeclareAcronym{tee}{
  short = TEE,
  long  = Trusted Execution Environment
}

\DeclareAcronym{cc}{
  short = CC,
  long  = Common Criteria
}

\DeclareAcronym{hsm}{
  short = HSM,
  long  = hardware security module
}

\DeclareAcronym{defi}{
  short = DeFi,
  long  = decentralised finance
}

\DeclareAcronym{aead}{
  short = AEAD,
  long  = authenticated encryption with associated data
}

\DeclareAcronym{kdf}{
  short = KDF,
  long  = key derivation function
}

\DeclareAcronym{pbkdf}{
  short = PBKDF,
  long  = password-based key derivation function
}

\DeclareAcronym{pbkdf2}{
  short = PBKDF2,
  long  = password-based key derivation function 2
}

\DeclareAcronym{kek}{
  short = KEK,
  long  = key encryption key
}

\DeclareAcronym{mfkdf}{
  short = MFKDF,
  long  = multi-factor key derivation function
}

\DeclareAcronym{dex}{
  short = DEX,
  long  = decentralised exchange
}

\DeclareAcronym{mcu}{
  short = MCU,
  long  = microcontroller unit
}

\DeclareAcronym{se}{
  short = SE,
  long  = secure element
}

\DeclareAcronym{os}{
  short = OS,
  long  = operating system
}

\DeclareAcronym{api}{
  short = API,
  long  = application programming interface
}

\DeclareAcronym{rpc}{
  short = RPC,
  long  = remote procedure call
}

\DeclareAcronym{dfd}{
  short = DFD,
  long  = data-flow diagram
}

\DeclareAcronym{ids}{
  short = IDS,
  long  = intrusion detection system
}

\DeclareAcronym{dkg}{
  short = DKG,
  long  = distributed key generation
}

\DeclareAcronym{mfa}{
  short = MFA,
  long  = multi-factor authentication
}

\DeclareAcronym{2fa}{
  short = 2FA,
  long  = two-factor authentication
}

\DeclareAcronym{ram}{
  short = RAM,
  long  = random-access memory
}

\DeclareAcronym{dram}{
  short = DRAM,
  long  = dynamic random-access memory
}

\DeclareAcronym{sram}{
  short = SRAM,
  long  = static random-access memory
}

\DeclareAcronym{rng}{
  short = RNG,
  long  = random number generator
}

\DeclareAcronym{ap}{
  short = AP,
  long  = access point
}

\DeclareAcronym{ui}{
  short = UI,
  long  = user interface
}

\DeclareAcronym{sca}{
  short = SCA,
  long  = side-channel analysis
}

\DeclareAcronym{see}{
  short = SEE,
  long  = secure execution environment
}

\DeclareAcronym{nee}{
  short = NEE,
  long  = non-secure execution environment
}

\DeclareAcronym{cnn}{
  short = CNN,
  long  = convolutional neural network
}

\DeclareAcronym{ec}{
  short = EC,
  long  = elliptic curve
}

\DeclareAcronym{mpc}{
  short = MPC,
  long  = multi-party computation
}

\DeclareAcronym{dao}{
  short = DAO,
  long  = decentralised autonomous organisation
}

\DeclareAcronym{hd}{
  short = HD,
  long  = Hierarchical Deterministic
}

\DeclareAcronym{mitm}{
  short = MITM,
  long  = man-in-the-middle
}

\DeclareAcronym{dos}{
  short = DoS,
  long  = denial-of-service
}

\DeclareAcronym{ddos}{
  short = DDoS,
  long  = distributed denial-of-service
}

\DeclareAcronym{syn}{
  short = SYN,
  long  = synchronise
}

\DeclareAcronym{adb}{
  short = ADB,
  long  = Android Debug Bridge
}

\DeclareAcronym{mac}{
  short = MAC,
  long  = Media Access Control
}

\DeclareAcronym{puf}{
  short = PUF,
  short-plural = PUFs,
  long  = physically unclonable function,
  long-plural = physically unclonable functions
}

\DeclareAcronym{utxo}{
  short = UTXO,
  short-plural = UTXOs,
  long  = unspent transaction output,
  long-plural = unspent transaction outputs
}

\DeclareAcronym{tls}{
  short = TLS,
  long  = Transport Layer Security
}

\DeclareAcronym{nvd}{
  short = NVD,
  long  = National Vulnerability Database
}

\DeclareAcronym{cfi}{
  short = CFI,
  long  = Control Flow Integrity
}

\begin{document}
\let\WriteBookmarks\relax
\def\floatpagepagefraction{1}
\def\textpagefraction{.001}



\shorttitle{SoK: Design, Vulnerabilities, and Security Measures of Cryptocurrency Wallets}   

\shortauthors{Erinle et~al.}

\title [mode = title]{SoK: Design, Vulnerabilities, and Security Measures of Cryptocurrency Wallets} 

\author[1,2]{Yimika Erinle}
[
                        orcid=0009-0003-9151-9114          
                        ]

\credit{Conceptualization of this study, Methodology, Software}

\affiliation[1]{organization={
DLT Science Foundation},
              citysep={}, 
                postcode={WC2H 2JQ}, 
                state={London},
                country={United Kingdom}
                }
                
\author[3]{Yathin Kethepalli}
[
                        orcid=0000-0002-1018-5251          
                        ]

\author[4]{Yebo Feng}
[
                        orcid=0000-0002-7235-2377          
                        ]


\credit{Data curation, Writing - Original draft preparation}

\affiliation[2]{organization={Centre for Blockchain Technologies, University College London},
                postcode={WCIE 6BT}, 
                postcodesep={}, 
                city={London},
                country={United Kingdom}      
                }

\author[1,2]{Jiahua Xu}
[
                        orcid=0000-0002-3993-5263          
                        ]


\affiliation[3]{organization={GlueX Protocol},
                postcode={2595VG}, 
                state={The Hague}, 
                country={Netherlands}
                }

\affiliation[4]{organization={Nanyang Technological University},
                postcode={639798}, 
                country={Singapore}
                }








\begin{abstract}
With the advent of decentralised digital currencies powered by blockchain technology, a new era of peer-to-peer transactions has commenced. The rapid growth of the cryptocurrency economy has led to the increased use of transaction-enabling wallets, making them a focal point for security risks. As the frequency of wallet-related incidents rises, there is a critical need for a systematic approach to measure and evaluate these attacks, drawing lessons from past incidents to enhance wallet security.

In response, we introduce a multi-dimensional design taxonomy for legacy and emerging wallets. We classify existing industry wallets based on this taxonomy, identify previously occurring vulnerabilities and discuss the security implications of design decisions. We also systematise threats to the wallet mechanism and analyse the adversary's goals, capabilities and required knowledge. We present a multi-layered attack framework and investigate 85 incidents between 2012 and 2025, accounting for a total of \$6.98B. Following this, we classify defence implementations for these attacks on the precautionary and remedial axes. We map the mechanism and design decisions to vulnerabilities, attacks, and possible defence methods to discuss various insights. 



\end{abstract}


\maketitle



\section{Introduction}
\label{sec:introduction}

Pioneered by Bitcoin \cite{NakamotoBitcoin:System}, peer-to-peer transactions have evolved into a digital ecosystem of decentralised financial applications on the blockchain. Building on this foundation with self-executing smart contracts on blockchain networks such as Ethereum, \ac{defi} protocols enable decentralised lending \cite{arora2024secplf}, exchanges \cite{xu2023sok}, derivatives \cite{luo2025piercing}, insurance \cite{cousaert2022token}, and numerous other financial applications \cite{cousaert2022sok, mita2019stablecoin, luo2025llm}. As the user-facing component, wallets intricately trigger various transactions.

A wallet is a transaction-facilitating tool that manages user authentication to enable digital signing of transactions. It broadcasts these messages to a blockchain network to confirm their validity. When initiating a transaction, wallets use a private key to sign and broadcast the signature to the blockchain network \cite{khan2022gas}. Private key security is therefore critical, as incidents such as the Mt. Gox exchange attack (850,000 BTC) have resulted in significant financial losses for individual users and entities relying on the service \cite{mtgox_hack}. Additional attacks on KuCoin \cite{kucoinNew}, Vulcan Forged \cite{VulcanHack}, Infarno \cite{infarno}, WazirX \cite{Explained:2024g}, and ByBit \cite{bybit_certik} have demonstrated that both custodial and non-custodial wallets present attractive targets.

This paper introduces a novel multi-dimensional cryptocurrency wallet taxonomy that extends beyond earlier approaches by covering both legacy and emerging wallets. The taxonomy reveals how specific design decisions correlate with known threat occurrences (\autoref{sec:wallet-taxonomy}). We systematise threats (\autoref{sec:threat_framework}) and attacks (\autoref{sec:attack-framework}), which enables us to suggest potential defence strategies (\autoref{sec:defense-strategies}). We then discuss our analysis of design elements, attack vectors, and defence types in \autoref{sec:insights}. In summary, our contributions are as follows:
\begin{itemize}
\item \textbf{Wallet Design Taxonomy:} We provide a taxonomy to analyse the design of various existing wallet types and propose new wallet designs. We also outline the threats to existing wallet designs based on our threat model.
\item \textbf{Wallet Attacks Framework:} We systematise and analyse various attack methods, techniques and targets in literature. We then analyse 85 notable wallet incidents between 2012 and 2025 and investigate the attack gaps between academia and industry.
\item \textbf{Defence Strategies:} We recommend defence methods based on the overall mitigation approach, incorporating both proactive and reactive approaches. We also analyse the influence of defence methods in mitigating attacks.
\end{itemize}

To facilitate independent verification, all datasets and code used in this study are publicly available.\footnote{GitHub repository at \url{https://github.com/xujiahuayz/crypto-wallets}.}

\section{Related Works}
\label{sec:related-work}

\subsection{Key Management}
\label{sec:key-management}

Several studies have explored key management mechanisms. Courtois and Mercer \cite{Courtois2017StealthSystems} compare key management solutions with a focus on stealth addresses. Mangipudi et al. \cite{Mangipudi2023UncoveringCrypto-Wallets} investigate key management from the wallet users' perspective. He et al. \cite{He2018AScheme} propose a secure key management scheme based on semi-trusted social networks. Di Angelo and Salzer \cite{di2020characteristics} analyse the functionality of smart contracts for key management through transaction data. Most recently, Chatzigiannis et al. \cite{chaganti2022comprehensive} propose a framework that formally evaluates hybrid recovery setups, highlighting key-management choices. Our study adopts a threat-centric view, mapping each key management technique in our multi-dimensional design taxonomy to specific attacks.

\subsection{Wallet Taxonomy}
\label{sec:wallet-taxonomy}

Prior research has proposed various methods to classify key management mechanisms \cite{bonneau2015sok, eskandari2018first, karantias2020sok, Homoliak2024SoK:Factors}.
Early wallet taxonomies by Bonneau et al. \cite{bonneau2015sok} and Eskandari et al. \cite{eskandari2018first} survey key management techniques such as password-protected files, paper-based methods, \acf{hsm} systems, password-derived wallets, and hosted services. However, this classification was confined to a single axis of key storage. Karantias \cite{karantias2020sok} contributes a protocol-centric taxonomy, examining light, full, and superlight clients and evaluating performance and security trade-offs. However, this approach does not extend to design elements such as key recovery methods or smart contract wallets, and lacks a mapping to threats and attack methods. Homoliak et al. \cite{Homoliak2024SoK:Factors} introduce an authentication-focused classification, examining k-factor and threshold-based co-signing solutions. While this emphasises the importance of multi-factor authentication in wallets, it only examines one of the several design elements we analyse.

By contrast, our taxonomy unifies multiple design dimensions into one integrative framework. These dimensions include custody model, key distribution, infrastructure (software or hardware), authentication, authorisation policies, and user recovery mechanisms. We include hardware wallets, exchange-based custodial solutions, shared-custodial implementations, non-custodial wallets, \acf{mpc} wallets, and smart contract wallets in a consistent scheme. This approach bridges the gap between academic and industry viewpoints.

\subsection{Wallet Attack and Security}
\label{sec:wallet-security}

A broad line of work surveys blockchain vulnerabilities and defences \cite{Li2020ASystems, Guo2022ASecurity, Chen2020ADefenses, zhou2023sok}. Chen et al. \cite{Chen2020ADefenses} focus on Ethereum’s protocol-layer issues. Researchers also analyse specific wallet mechanisms; in particular, HSM-focused defence studies \cite{Shbair2021HSM-basedBlockchain, Gotte2021TechAttacks}. Additional studies investigate specific vectors such as phishing \cite{andryukhin2019phishing} and desktop-wallet RPC pitfalls \cite{bui2019pitfalls}. Others scope security across wallet types \cite{Das2019AWallets}, access key management impacts \cite{Eyal2022OnDesign}, and review attacks and defences in academia \cite{Houy2023}.

Our work differs by adopting a multi-layered defence perspective and incorporating real-world incident analysis to evaluate how design choices influence attacks. This approach bridges academic models with industry practice.

\begin{table}[!ht]
  \centering
  \renewcommand{\arraystretch}{1.2}
  \resizebox{\linewidth}{!}{%
    \begin{tabular}{r*{6}{c}*{4}{c}*{4}{c}}
      \toprule
      & \multicolumn{6}{c}{\textbf{Subjects Covered}}
      & \multicolumn{4}{c}{\textbf{Methodology}}
      & \multicolumn{4}{c}{\textbf{Scope}}
      \\
      \cmidrule(lr){2-7} \cmidrule(lr){8-11} \cmidrule(lr){12-15}
      \textbf{Reference}
      & \rot[90]{Key Cryptography}
      & \rot[90]{Key Management}
      & \rot[90]{Key Recovery}
      & \rot[90]{Attack Methods}
      & \rot[90]{Security Measures}
      & \rot[90]{Privacy Techniques}
      & \rot[90]{Literature}
      & \rot[90]{Taxonomisation}
      & \rot[90]{Analysis}
      & \rot[90]{Case Study}
      & \rot[90]{Wallet Software}
      & \rot[90]{Wallet Hardware}
      & \rot[90]{Smart Contract Wallet}
      & \rot[90]{Blockchain Network} \\
      \midrule
      This Study
      & \CIRCLE & \CIRCLE & \CIRCLE & \CIRCLE & \CIRCLE & \Circle
      & \CIRCLE & \CIRCLE & \CIRCLE & \CIRCLE
      & \CIRCLE  & \CIRCLE  & \CIRCLE  & \Circle \\
      \cite{bonneau2015sok}
      & \CIRCLE & \CIRCLE & \Circle  & \Circle  & \CIRCLE & \CIRCLE
      & \CIRCLE & \CIRCLE & \CIRCLE & \Circle
      & \CIRCLE  & \Circle   & \Circle   & \CIRCLE \\
      \cite{eskandari2018first}
      & \Circle   & \CIRCLE  & \CIRCLE  & \Circle   & \CIRCLE  & \Circle
      & \CIRCLE  & \CIRCLE  & \CIRCLE  & \Circle
      & \CIRCLE  & \Circle   & \Circle   & \Circle \\
      \cite{karantias2020sok}
      & \Circle   & \Circle   & \Circle   & \Circle   & \CIRCLE  & \CIRCLE
      & \CIRCLE  & \CIRCLE  & \CIRCLE  & \Circle
      & \CIRCLE  & \Circle   & \Circle   & \Circle \\
      \cite{Homoliak2020SmartOTPs:Wallets}
      & \Circle   & \CIRCLE  & \Circle   & \CIRCLE  & \CIRCLE  & \Circle
      & \CIRCLE  & \CIRCLE  & \CIRCLE  & \Circle
      & \CIRCLE  & \CIRCLE  & \Circle   & \CIRCLE \\
      \cite{Houy2023}
      & \Circle   & \CIRCLE  & \CIRCLE  & \CIRCLE  & \CIRCLE  & \CIRCLE
      & \CIRCLE  & \CIRCLE  & \CIRCLE  & \Circle
      & \CIRCLE  & \CIRCLE  & \Circle   & \CIRCLE \\
      \cite{suratkar2020cryptocurrency}
      & \CIRCLE & \CIRCLE & \CIRCLE & \Circle  & \Circle  & \Circle
      & \CIRCLE & \CIRCLE & \Circle   & \Circle
      & \CIRCLE  & \Circle   & \Circle   & \Circle \\
      \cite{bui2019pitfalls}
      & \Circle  & \Circle  & \Circle  & \CIRCLE & \CIRCLE & \Circle
      & \CIRCLE & \CIRCLE & \CIRCLE  & \Circle
      & \CIRCLE  & \Circle   & \Circle   & \Circle \\
      \cite{zaghloul2020bitcoin}
      & \Circle  & \Circle  & \Circle  & \Circle  & \CIRCLE & \CIRCLE
      & \CIRCLE & \CIRCLE & \CIRCLE  & \Circle
      & \CIRCLE  & \Circle   & \Circle   & \CIRCLE \\
      \cite{li2020android}
      & \Circle  & \Circle  & \Circle  & \CIRCLE & \CIRCLE & \Circle
      & \CIRCLE & \Circle   & \CIRCLE  & \Circle
      & \CIRCLE  & \Circle   & \Circle   & \Circle \\
      \cite{Dai2018SBLWT:Trustzone}
      & \CIRCLE & \CIRCLE & \Circle  & \CIRCLE & \CIRCLE & \Circle
      & \Circle  & \CIRCLE  & \CIRCLE  & \Circle
      & \CIRCLE  & \Circle   & \Circle   & \Circle \\
      \cite{volety2019cracking}
      & \Circle  & \Circle  & \Circle  & \CIRCLE & \CIRCLE & \Circle
      & \CIRCLE & \Circle   & \CIRCLE  & \Circle
      & \CIRCLE  & \Circle   & \Circle   & \Circle \\
      \cite{8966739}
      & \CIRCLE & \CIRCLE & \CIRCLE & \CIRCLE & \CIRCLE & \CIRCLE
      & \CIRCLE & \Circle   & \CIRCLE  & \Circle
      & \Circle   & \CIRCLE   & \Circle   & \Circle \\
      \cite{rezaeighaleh2020improving}
      & \CIRCLE & \CIRCLE & \CIRCLE & \Circle  & \CIRCLE & \Circle
      & \CIRCLE & \CIRCLE & \CIRCLE  & \Circle
      & \Circle   & \CIRCLE   & \Circle   & \Circle \\
      \cite{Urien2021InnovativeWallets}
      & \Circle  & \Circle  & \Circle  & \CIRCLE & \CIRCLE & \Circle
      & \CIRCLE & \CIRCLE & \Circle   & \Circle
      & \Circle   & \CIRCLE   & \Circle   & \Circle \\
      \cite{Rezaeighaleh2020MultilayeredWallet}
      & \Circle  & \Circle  & \CIRCLE & \Circle  & \Circle  & \CIRCLE
      & \CIRCLE & \CIRCLE & \Circle   & \Circle
      & \Circle   & \CIRCLE   & \Circle   & \Circle \\
      \cite{rezaeighaleh2019new}
      & \CIRCLE & \CIRCLE & \CIRCLE & \Circle  & \CIRCLE & \Circle
      & \CIRCLE & \Circle   & \CIRCLE  & \Circle
      & \Circle   & \CIRCLE   & \Circle   & \Circle \\
      \cite{di2020characteristics}
      & \Circle  & \Circle  & \Circle  & \Circle  & \Circle  & \Circle
      & \CIRCLE & \CIRCLE & \CIRCLE  & \Circle
      & \Circle   & \Circle    & \CIRCLE   & \Circle \\
\cite{Homoliak2024SoK:Factors}
& \Circle & \CIRCLE & \Circle & \Circle & \CIRCLE & \Circle
& \CIRCLE & \CIRCLE & \CIRCLE & \Circle
& \CIRCLE & \CIRCLE & \Circle & \Circle \\

\cite{andryukhin2019phishing}
& \Circle & \Circle & \Circle & \CIRCLE & \CIRCLE & \Circle
& \CIRCLE & \Circle & \CIRCLE & \Circle
& \Circle & \Circle & \Circle & \Circle \\

\cite{Chen2020ADefenses}
& \Circle & \Circle & \Circle & \CIRCLE & \CIRCLE & \Circle
& \CIRCLE & \CIRCLE & \CIRCLE & \Circle
& \Circle & \Circle & \Circle & \CIRCLE \\

\cite{Courtois2017StealthSystems}
& \CIRCLE & \CIRCLE & \Circle & \CIRCLE & \CIRCLE & \CIRCLE
& \CIRCLE & \Circle & \CIRCLE & \Circle
& \CIRCLE & \Circle & \Circle & \CIRCLE \\

\cite{Das2019AWallets}
& \CIRCLE & \CIRCLE & \Circle & \Circle & \CIRCLE & \Circle
& \Circle & \Circle & \CIRCLE & \Circle
& \CIRCLE & \Circle & \Circle & \Circle \\

\cite{Eyal2022OnDesign}
& \Circle & \CIRCLE & \Circle & \Circle & \CIRCLE & \Circle
& \Circle & \CIRCLE & \CIRCLE & \Circle
& \CIRCLE & \CIRCLE & \Circle & \Circle \\

\cite{Gotte2021TechAttacks}
& \CIRCLE & \CIRCLE & \Circle & \CIRCLE & \CIRCLE & \Circle
& \Circle & \Circle & \CIRCLE & \Circle
& \Circle & \CIRCLE & \Circle & \Circle \\

\cite{Guo2022ASecurity}
& \Circle & \CIRCLE & \Circle & \CIRCLE & \CIRCLE & \Circle
& \CIRCLE & \CIRCLE & \CIRCLE & \Circle
& \CIRCLE & \CIRCLE & \Circle & \CIRCLE \\

\cite{He2018AScheme}
& \Circle & \CIRCLE & \CIRCLE & \Circle & \CIRCLE & \Circle
& \Circle & \Circle & \CIRCLE & \Circle
& \CIRCLE & \Circle & \Circle & \Circle \\

\cite{Houy2023}
& \CIRCLE & \CIRCLE & \CIRCLE & \CIRCLE & \CIRCLE & \Circle
& \CIRCLE & \CIRCLE & \CIRCLE & \Circle
& \CIRCLE & \CIRCLE & \Circle & \CIRCLE \\

\cite{Li2020ASystems}
& \Circle & \CIRCLE & \Circle & \CIRCLE & \CIRCLE & \Circle
& \Circle & \Circle & \CIRCLE & \Circle
& \CIRCLE & \Circle & \Circle & \Circle \\

\cite{Mangipudi2023UncoveringCrypto-Wallets}
& \Circle & \CIRCLE & \Circle & \Circle & \Circle & \Circle
& \Circle & \Circle & \CIRCLE & \CIRCLE
& \CIRCLE & \Circle & \Circle & \Circle \\

\cite{Shbair2021HSM-basedBlockchain}
& \CIRCLE & \CIRCLE & \Circle & \CIRCLE & \CIRCLE & \Circle
& \Circle & \Circle & \CIRCLE & \CIRCLE
& \Circle & \CIRCLE & \Circle & \Circle \\

\cite{zhou2023sok}
& \Circle & \Circle & \Circle & \CIRCLE & \CIRCLE & \Circle
& \CIRCLE & \CIRCLE & \CIRCLE & \Circle
& \Circle & \Circle & \CIRCLE & \CIRCLE \\

\cite{chatzigiannis2025composability}
& \Circle & \CIRCLE & \CIRCLE & \CIRCLE & \CIRCLE & \Circle
& \Circle & \Circle & \CIRCLE & \CIRCLE
& \CIRCLE & \Circle & \CIRCLE & \Circle \\

      \bottomrule
    \end{tabular}%
  }
  \caption{Overview of related works. (\CIRCLE: include, \Circle: not include)}
  \label{Literature-Gap-Table-1}
\end{table}

\subsection{Addressing Literature Gaps}
\label{sec:gaps-in-literature}

Despite various studies on specific wallet types, mechanisms, and attack vectors, there is a lack of comprehensive examination spanning wallet design taxonomy, attack methods, incident analysis, security measures, and case studies, as shown in \autoref{Literature-Gap-Table-1}. Moreover, our design taxonomy is mapped with a detailed threat model and defence strategies, allowing a systematic evaluation of each design's security trade-offs. This comprehensive coverage and empirical attack data distinguish our work from prior classification-focused surveys. Our study bridges this gap, providing a holistic understanding crucial for advancing wallet security.

\section{Generalised Wallet Mechanism}
\label{sec:wallet_mechanism}

Cryptocurrency wallets facilitate state transitions by securely managing cryptographic keys and authorising transaction execution on the blockchain. To analyse wallet design and security, we first define a wallet. This definition underpins our mechanism, taxonomy, threat model, attack taxonomy, and security measures.

\begin{definition}[Cryptocurrency Wallet]
A wallet is a system that typically generates a private key, also known as the secret key (\textcolor{teal}{\textit{sk}}), and securely stores it in an encrypted form (\textcolor{teal}{\textit{enc\_sk}}), enabling an authenticated owner to sign transactions that are broadcast to the blockchain.
\end{definition}

\subsection{Key Generation}
\label{sec:key_generation}

\begin{figure}[!b]
	\centering \includegraphics[scale=0.85]
 {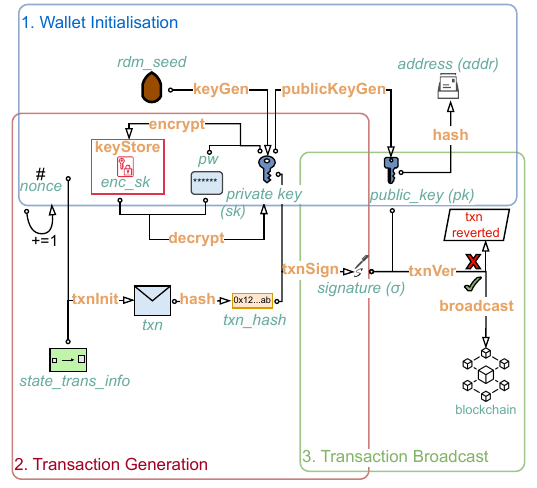}
	\caption{Generalised cryptocurrency wallet mechanism showing \hyperref[algo:cryptocurrency-wallet]{Algorithm 1}, \hyperref[algo:transaction-signing]{Algorithm 2} and \hyperref[algo:transaction-broadcast]{Algorithm 3}.}
	\label{fig:wallet-mechanism}
\end{figure}

The wallet initialisation process, detailed in \hyperref[algo:cryptocurrency-wallet]{Algorithm 1}, specifies private key generation, public key generation, public address derivation and private key encryption for secure storage. As shown in \autoref{fig:wallet-mechanism}, the internal flow of the wallet begins with private key (\teal{\textit{$sk$}}) generation from a random seed (\teal{\textit{$rdm\_seed$}}). The corresponding public key (\textcolor{teal}{\textit{pk}}) is then derived from \textcolor{teal}{\textit{sk}} using the signature scheme and curve required by the target chain. Bitcoin, Ethereum, and Avalanche\footnote{Avalanche's C-Chain is \acf{evm} compatible and therefore inherits \textit{secp256k1}. Hedera introduced optional \textit{secp256k1} accounts in 2023 for \acs{evm} compatibility; however, \textit{ed25519} remains the default.} all rely on the \textit{\acf{ecdsa}} over the \textit{secp256k1} curve by default \cite{sec2}. Solana and Hedera default to the \textit{\acf{eddsa}} curve \textit{ed25519} \cite{rfc8032}, whereas the XRP Ledger supports both \textit{\acs{ecdsa}}/\textit{secp256k1} and \textit{\acs{eddsa}}/\textit{ed25519}.

Once the key pair is generated and \textcolor{teal}{\textit{pk}} is obtained, the wallet hashes \textcolor{teal}{\textit{pk}} to produce the address (\textcolor{teal}{\textit{addr}}). Users share this address to receive funds. In account-based blockchains, the wallet queries \textcolor{teal}{\textit{addr}} via an \acf{rpc} to fetch the current nonce (\textcolor{teal}{\textit{nonce}}). The nonce is initialised to 0 and preserves the sequential order of outgoing transactions.

Beyond curve selection, contemporary wallet software adheres to a concise suite of public standards. \acf{bip} 32 \cite{bip32} and \acf{slip} 10 \cite{slip10} define \acf{hd} key derivation for \textit{secp256k1} and \textit{ed25519} curves, respectively. 

Mnemonic phrases, as defined in \acs{bip}-39 \cite{bip39_}, are the widely adopted standard for representing seeds in a human-readable form. \acs{slip}-39 \cite{slip39} extends this by applying Shamir’s Secret Sharing to mnemonic phrases, enabling distributed or threshold-based recovery of wallet seeds. These mechanisms enable secure \textcolor{teal}{\textit{sk}} recovery in case of device loss or failure (see \autoref{sec:design-rec}). At the account-level, wallet use standard derivation paths such as \acs{bip}-44 \cite{bip44}, \acs{bip}-49 \cite{bip49} and \acs{bip}-84 \cite{bip84} to deterministically derive multiple accounts and address types from a single seed.

\begin{algorithm}
\caption{Wallet initialisation}
\label{algo:cryptocurrency-wallet}
\begin{algorithmic}[1]
    \State \textbf{Input:} \textcolor{teal}{\textit{rdm\_seed}}: \olive{bin}, \textcolor{teal}{\textit{pw}}: \olive{str}
    \State \textcolor{teal}{\textit{sk}} = \orange{{\textit{keyGen}}}(\textcolor{teal}{\textit{rdm\_seed}})
    \State \textcolor{teal}{\textit{pk}} = \orange{\textit{publicKeyGen}}(\textcolor{teal}{\textit{sk}})
    \State \textcolor{teal}{\textit{enc\_sk}} = \orange{\textit{encrypt}}(\textcolor{teal}{\textit{sk}}, \textcolor{teal}{\textit{pw}})
    \State \textcolor{teal}{\textit{addr}} = {\orange{\textit{hash}}}(\textcolor{teal}{\textit{pk}})
    \State \textcolor{teal}{\textit{nonce}} = 0
\end{algorithmic}
\end{algorithm}

\subsection{Key Storage}
\label{sec:key-storage}

Following its generation, \textcolor{teal}{\textit{sk}} is stored and encrypted using a \acf{kek} that we refer to simply as the password (\textcolor{teal}{\textit{pw}}), as shown in \hyperref[algo:cryptocurrency-wallet]{Algorithm 1}. In practice, \textcolor{teal}{\textit{pw}} is an abstract input that may be a traditional text password, a numeric PIN, a device-derived biometric secret, or a composite value obtained through multi-factor authentication. The ensuing \acf{kdf} output serves as the \acs{kek}. \acs{kek}s are typically derived with a \acf{pbkdf}, such as \textit{PBKDF2-\acs{hmac}-\acs{sha}-256} \cite{rfc8018}, \textit{scrypt} \cite{rfc7914}, or the memory-hard \textit{Argon2id} \cite{rfc9106}. The resulting \acs{kek} then protects \textcolor{teal}{\textit{sk}} under an \acf{aead} cipher such as \textit{\acs{aes}-256-\acs{gcm}}\,\cite{fips197} or \textit{XChaCha20\allowbreak -Poly1305} \cite{rfc8439}. The encrypted private key (\textcolor{teal}{\textit{enc\_sk}}) remains secure, with \textcolor{teal}{\textit{pw}} required for both decryption and transaction signing. Secure \textcolor{teal}{\textit{sk}} storage is governed by the interplay of several factors described in \autoref{sec:wallet-taxonomy}.

\subsection{Transaction Management}
\label{sec:transaction_management}

\begin{definition}[Transaction]
\label{sec:def-trans}
A transaction (\textcolor{teal}{\textit{txn}}) is a structured message created by a wallet that enables state change executions on the blockchain. These state changes include token transfers and smart contract interactions.
\end{definition}

\subsubsection{Transaction Generation}
\label{sec:transaction_signing}

Transaction generation begins with creating the transaction message (\textcolor{teal}{\textit{txn}}) by inputting the state transition information (\textcolor{teal}{\textit{state\_trans\_info}}). The message (\textcolor{teal}{\textit{txn}}) is then hashed to produce the transaction hash (\textcolor{teal}{\textit{txn\_hash}}). Following transaction creation, the sender signs the transaction and provides \textcolor{teal}{\textit{pw}} to decrypt the private key (\textcolor{teal}{\textit{sk}}). The signing algorithm takes the decrypted private key (\textcolor{teal}{\textit{sk}}) and \textcolor{teal}{\textit{txn\_hash}} as inputs to generate the signature (\teal{$\sigma$}), which authorises the transaction (see \hyperref[algo:transaction-signing]{Algorithm 2}).

\subsubsection{Transaction Broadcast}
\label{sec:transaction_broadcast}

The signature (\teal{$\sigma$}) is verified using the sender's public key (\textcolor{teal}{\textit{pk}}) to assert its validity, as shown in \hyperref[algo:transaction-broadcast]{Algorithm 3}. If \teal{$\sigma$} is invalid, the transaction is rejected and not processed further. Conversely, if \teal{$\sigma$} is valid, the transaction is broadcast to the blockchain. 

\begin{algorithm}[!b]
    \caption{Transaction generation}
    \label{algo:transaction-signing}
    \begin{algorithmic}[1]  
        \State \textbf{Input:} \textcolor{teal}{\textit{nonce}}: \textcolor{olive}{int}, \textcolor{teal}{\textit{state\_trans\_info}} : \textcolor{olive}{str}, \textcolor{teal}{\textit{enc\_sk}}: \textcolor{olive}{bytes}, \textcolor{teal}{\textit{pw}}: \textcolor{olive}{str}
        \State \textbf{Output:} \textcolor{teal}{$\sigma$}: \textcolor{olive}{bytes}

        \State \textcolor{teal}{\textit{nonce}} += 1
        \State \textcolor{teal}{\textit{txn}} = \textcolor{orange}{\textit{txnInit}}(\textcolor{teal}{\textit{state\_trans\_info}}, \textcolor{teal}{\textit{nonce}})
        \State \textcolor{teal}{\textit{txn\_hash}} = \textcolor{orange}{\textit{hash}}(\textcolor{teal}{\textit{txn}})
        \State \textcolor{teal}{\textit{sk}} = \textcolor{orange}{\textit{decrypt}}(\textcolor{teal}{\textit{enc\_sk}}, \textcolor{teal}{\textit{pw}})
        \State \textcolor{teal}{$\sigma$} = \textcolor{orange}{\textit{txnSign}}(\textcolor{teal}{\textit{txn\_hash}}, \textcolor{teal}{\textit{sk}})

        \State \textbf{return:} \textcolor{teal}{$\sigma$}  
    \end{algorithmic}
\end{algorithm}

\section{Methodology}
\label{sec:methodology}

Our methodology systematically bridges academic research and industry practice by analysing cryptocurrency wallet security across four axes: design, vulnerabilities, attacks, and defence measures.


\subsection{Procedure}

\subsubsection{Design Taxonomy and Vulnerability}

Our wallet design survey is structured as follows. We first perform reverse-engineering of specific vulnerable wallets to map vulnerabilities explicitly to underlying design features. Unique wallet features relevant to security were carefully documented and compared across wallet categories. The results of this analysis are summarised systematically in our wallet taxonomy table (\autoref{tab:wlt._taxonomy}), enabling structured comparisons and insight into security-usability trade-offs.

\subsubsection{Attack Methods}

Following our design and threat analysis, we examine wallet attack methods in both academia and industry through a three-phase process. First, we conduct a comprehensive review of academic literature and industry incidents. We examine 33 peer-reviewed papers alongside 85 real-world incidents (2012–2025) documented in grey literature sources such as \href{https://rekt.news/}{Rekt News} and \href{https://www.slowmist.com/}{Slowmist}.

To expand the reviewed literature scope, we conduct forward and backward reference searches. Following this, we categorise attacks using a three-tier framework to establish clarity and consistency. Attacks are classified hierarchically by their mechanism-centric goal (e.g. bypass the authentication mechanism), method (e.g., credential cracking), and vector (e.g., dictionary attack). We analyse industry incidents and identify patterns related to our design taxonomy or attack categorisation. Lastly, we perform a gap analysis to evaluate the alignment between academic research and industry practices.

\subsubsection{Security Measures}

Our security measures analysis begins by identifying proposed and implemented defensive strategies documented within the 33 academic papers focused on attack methods. We employ forward and backward reference searches to expand the scope of our reviewed literature to 61 unique references, retrieving an additional 28 academic papers. In addition, we consult grey literature sources on security measures. Each security measure is mapped to an identified wallet attack vector and classified based on the approach (e.g. proactive or reactive). 

\subsubsection{Case Studies}
We conduct in-depth case studies to illustrate the practical application of our framework. We systematically select representative wallet incidents based on severity and distinctiveness. Each case study follows a structured approach: 
\begin{enumerate*}[label=(\arabic*)]
  \item describing the wallet’s design using our taxonomy, 
  \item detailing exploited vulnerabilities and threats, 
  \item outlining the adversary’s goals, capabilities, and attack sequences and 
  \item recommending security measures.
\end{enumerate*}
By integrating these real-world examples, we provide actionable insights into the interplay of wallet design, threats, and mitigation strategies.

\begin{algorithm}[!b]
    \caption{Transaction broadcast}
    \label{algo:transaction-broadcast}
    \begin{algorithmic}[1]  
        \State \textbf{Input:} \textcolor{teal}{$\sigma$}: \textcolor{olive}{bytes}, \textcolor{teal}{\textit{pk}}: \textcolor{olive}{hex}

        \State \textcolor{teal}{\textit{verified}} = \textcolor{orange}{\textit{txnVer}}(\textcolor{teal}{$\sigma$}, \textcolor{teal}{\textit{pk}})
        \State \textcolor{orange}{\textit{assert}}(\textcolor{teal}{\textit{verified}}, \enquote{transaction failed})
        \State \textcolor{orange}{\textit{broadcast}}(\textcolor{teal}{$\sigma$}, \textcolor{teal}{\textit{pk}})

    \end{algorithmic}
\end{algorithm}

\subsection{Data Sources}

We sourced design variation, vulnerability, attacks and defence methods data from the following:

\begin{itemize}
    \item \textbf{\acs{cve} Database:} We query the \href{https://www.cve.org/}{\acf{cve}} databases to retrieve previously identified wallet vulnerabilities.
    \item \textbf{Academic Papers:} We systematically retrieve academic papers, which serve as the primary data source for a range of wallet attack vectors and defence implementations.
    \item \textbf{Grey Literature:} We discover incidents on custodial and non-custodial wallets between 2012 and 2025, with most sources from \href{https://rekt.news/}{Rekt News} and \href{https://www.slowmist.com/}{Slowmist}. Grey literature is also employed to retrieve additional vulnerabilities and security measures.
\end{itemize}

\begin{figure*}[!t]
    \centering
    \includegraphics[width=1.75\columnwidth]{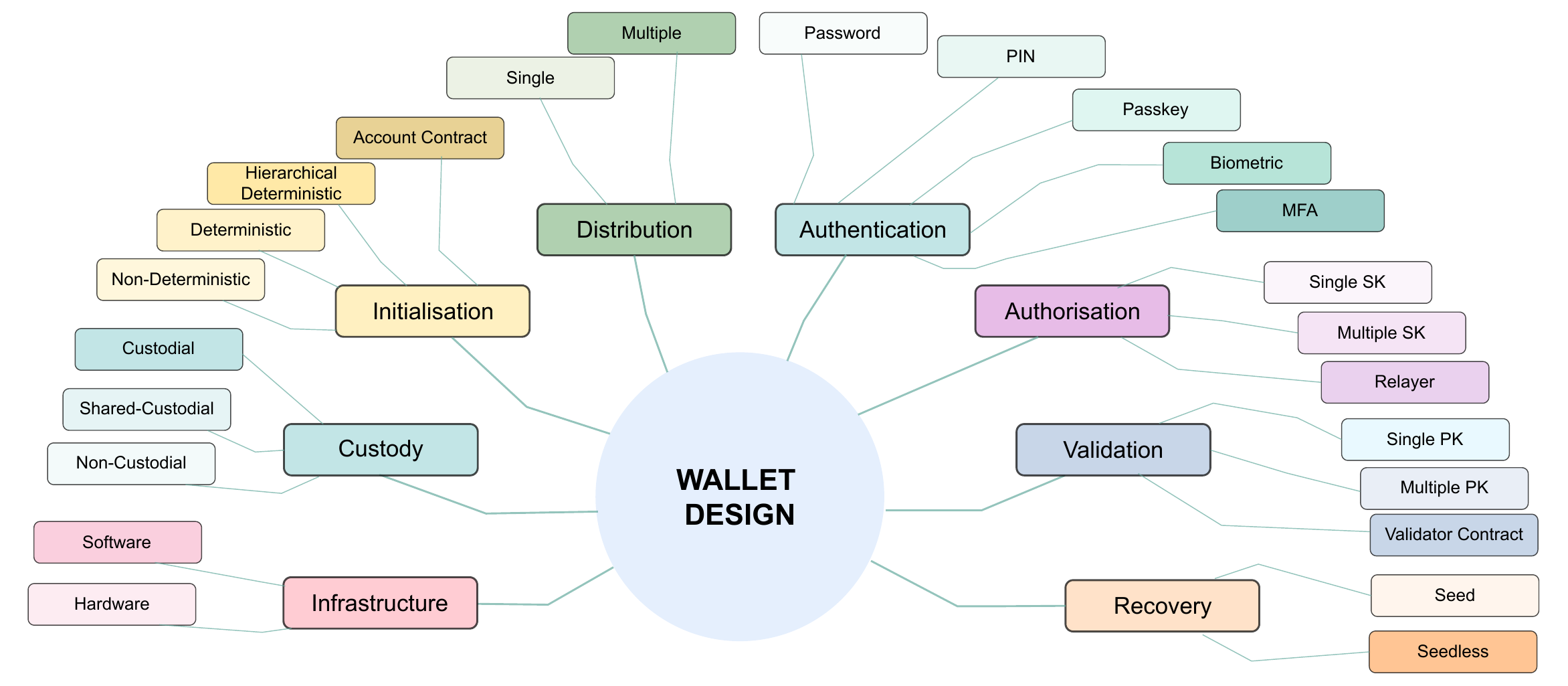}
    \caption{Multi-dimensional wallet design taxonomy for traditional and emerging wallets. \autoref{fig:wallet-taxonomy} maps industry wallets on three dimensions based on this taxonomy.}
    \label{fig:multi_axes}
\end{figure*}


\subsection{Inclusion Criteria}

Our resulting data conformed to the inclusion criteria below:

\begin{itemize}
    \item 
    \textbf{General Scope:} We limit our scope to exclude attacks on the blockchain protocol and on \ac{defi} protocols from our discussion or analysis.
    \item 
    \textbf{Vulnerability Inclusion:} We include wallet solutions with at least one \href{https://www.cve.org/}{\acs{cve}} or previously detected vulnerability from searches.
    \item \textbf{Design Inclusion:} We include wallets with previously identified vulnerabilities, as well as those with significant user bases (such as MetaMask, Trust Wallet) or \acf{aum} (centralised exchanges such as Coinbase Exchange, Binance Exchange) and wallets with novel features (Argent, Safe (previously Gnosis Safe), ZenGo and Ngrave).
    \item \textbf{Attack and Defence Inclusion:} We include only attack methods and defence implementations, which can be mapped to key components within the underlying mechanism.
    \item \textbf{Case Studies Inclusion:} We include two notable incidents exhibiting: one custodial breach with the largest recorded monetary loss and one non‑custodial compromise affecting the widest user base. These also provide comparative coverage of attacks against smart contract, hardware, and mobile wallet infrastructures.
\end{itemize}

\section{Wallet Design Taxonomy}
\label{sec:wallet-taxonomy}

We propose a design taxonomy for classifying and developing wallets that integrates traditional models and recent advances, as illustrated in \autoref{fig:multi_axes}. To develop this framework, we analyse various designs of wallets within the industry. We also identify known vulnerabilities and previous attacks associated with these wallets, as summarised in \autoref{tab:wlt._taxonomy}.


\subsection{Infrastructure}
\label{sec:infrastructure}

This design factor is centred on the private key (\textcolor{teal}{\textit{sk}}) or transaction management infrastructure (see \autoref{sec:wallet_mechanism}) the controlling entity employs.

\subsubsection{Software Wallets}
\label{sec:software-wallets} 

Software wallets are applications that manage private keys (\textcolor{teal}{\textit{sk}}) or transaction authorisation conditions within a software environment. Existing software infrastructure designs include desktop, browser, mobile and smart contract wallets, as demonstrated within \autoref{fig:wallet-taxonomy}. Desktop wallets are installed on computers and typically store \textcolor{teal}{\textit{enc\_sk}} in a local file within the computer's file system. Browser wallets present an alternative setup, with programs installed or built into the web browser and credentials are typically stored in the browser's local storage \cite{2024MetaMaskWallet}. Two existing designs are browser extensions, such as MetaMask and Phantom, and built-in browser-native wallets, such as Brave \cite{Brave2023BraveBrave}. 

Another prevalent wallet type is the mobile wallet, which is installed on devices with limited computing power and storage capability in comparison with PCs. Mobile wallets also typically store \textcolor{teal}{\textit{enc\_sk}} locally and can enhance security through mobile \acs{os} integrations such as the Android Keystore and iOS Keychain \cite{keystore}. However, if vulnerabilities are present in the operating system \autoref{sec:threat_class}, susceptibility to specific attacks that exploit these weaknesses exists (see \autoref{sec:privilege}).

To mitigate the risk of \textcolor{teal}{\textit{sk}} and \textcolor{teal}{\textit{rdm\_seed}} loss, smart contract wallets (e.g., Argent and Safe) are deployed on the blockchain to abstract typical \textcolor{teal}{\textit{sk}} management (see \autoref{sec:wallet_mechanism}) and create advanced transaction functions such as \acf{mfa}, ownership assignments, spending limits, and recovery mechanisms, often through integration with centralised or decentralised relayers \cite{di2020characteristics, erc4337}. 

TON Space, another smart contract wallet, allows users to create and sign transactions without leaving the chat by interacting through TON's standard Wallet-V4 account model \cite{tonwalletv4}. The key management functionality, bot-based transfers, and cloud backups are mediated through Telegram IDs and WebView sessions.  This approach shifts part of the trust boundary from the mobile operating system to Telegram’s \acs{api} and bot infrastructure, introducing centralisation risks \cite{beincrypto2025ton} and
exposing generic WebView attack surfaces \cite{halborn2024webview}. Despite their capabilities, smart contract wallets are susceptible to library vulnerabilities, implementation flaws, and access-control misconfigurations. These application logic vulnerabilities have resulted in significant financial losses in several cases \cite{palladino2017parity, Parisi2023WalletSecurity, bybit_certik}.

\subsubsection{Hardware Wallets}
\label{sec:hardware-wallets}

\begin{figure*}[!t]
    \includegraphics[scale=0.70]
 {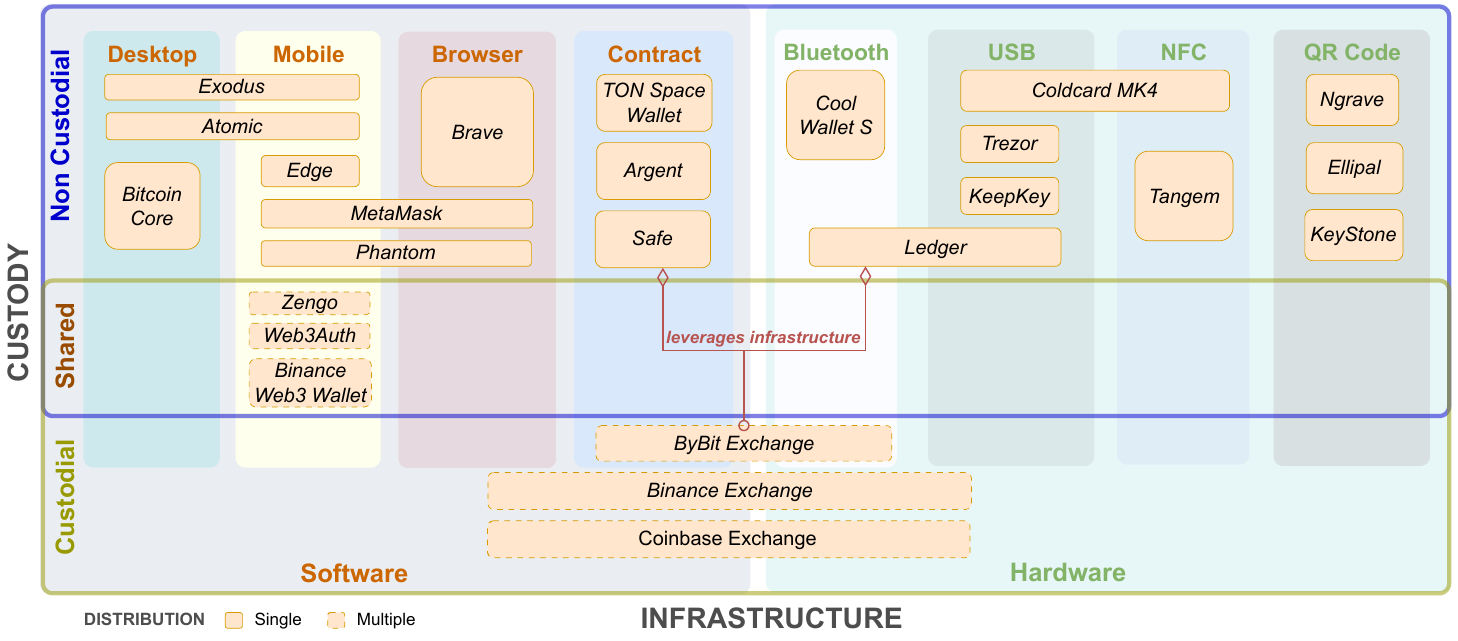}
	\caption{Wallet design taxonomy showing three of our eight dimensions (infrastructure, custody, and distribution), as detailed in \autoref{sec:infrastructure}, \autoref{sec:design-cust} and \autoref{sec:design-distr}.}
	\label{fig:wallet-taxonomy}
\end{figure*}

Hardware wallets typically involve \textcolor{teal}{\textit{sk}} management within a \acf{se} (e.g., microcontroller or smart card) to protect against tampering and facilitate the execution of cryptographic operations, such as transaction signing (see \autoref{sec:wallet_mechanism}). Isolated in design with no internet connectivity, their mechanism performs all cryptographic operations on an offline hardware device. They typically require a distinct online device to create and broadcast transactions \cite{ledgeracademy}. As shown in \autoref{fig:wallet-taxonomy}, the connection between both devices can be achieved through Bluetooth (e.g., Ledger), USB (e.g., Trezor), \acf{nfc} (e.g., Tangem) and QR codes (e.g., Ngrave). Specific hardware wallet vulnerabilities \cite{Cointelegraph2023LedgerRedefined, Ledger2018FirmwareFixed, CoinDesk2018SecurityAntennae, Freemindtronic2023LedgerHackers}, and attacks \cite{Akter2023AChallenges, hajdu2020using, wang2024efficient, courbon2016reverse} are discussed in \autoref{sec:vuln_phy} and \autoref{sec:physical-attacks}, respectively.

\subsection{Custody}
\label{sec:design-cust}

The degree of \textcolor{teal}{\textit{sk}} control by an entity or between one or more entities defines custody design. Custody setups include custodial, non-custodial and shared-custodial.

\subsubsection{Custodial}
\label{sec:custodial-wallets}

In this model, \textcolor{teal}{\textit{enc\_sk}} is stored by a trusted custodian (e.g., Coinbase Exchange, Binance Exchange, Kraken Exchange) who signs user-initiated transactions. The user relinquishes \textcolor{teal}{\textit{sk}} security to the custodian who fully controls the wallet operations (see \autoref{sec:wallet_mechanism}), while the user solely crafts transaction messages. Although most of the design factors for custodial wallets are not disclosed (see \autoref{tab:wlt._taxonomy}), a classification of their design can be conducted using our framework. In the table, we denote \enquote{\smallhalfcirc} representing user-facing infrastructure and \enquote{\smallfullcirc} the internal infrastructure employed by the custodian. 

Two notable design variations exist in custodial wallets. First, an omnibus setup aggregates and controls all users' funds under a few shared addresses, without a one-to-one correspondence between user accounts and addresses. Second, a segregated setup assigns each user a unique blockchain address, with the custodian retaining control of the associated private keys (\textcolor{teal}{\textit{sk}}) \cite{chalkias2022broken}.

\subsubsection{Non-Custodial}
\label{sec:non-custodial-wallets}

In non-custodial wallet architectures (e.g., MetaMask, Phantom, Ledger), the user does not relinquish control to any custodian party. Instead, a direct interaction between the user and the blockchain network exists in these setups with the user in full control of \textcolor{teal}{\textit{sk}}, to facilitate all the wallet operations (see \autoref{sec:wallet_mechanism}). With full autonomy, the user is solely responsible for securing \textcolor{teal}{\textit{sk}} and is more susceptible to insecure user interaction threats as well as other vulnerabilities (see \autoref{sec:threat_class}) and attacks such as social engineering attacks and malware-based attacks (see \autoref{sec:application-attacks}) which aim to exploit user negligence. While non-custodial wallets are expected not to have credential control, a few incidents in the past (e.g., Slope Wallet \cite{CoinTelegraph}) have resulted in \textcolor{teal}{\textit{sk}} compromise due to poor implementation practices, insecure storage of sensitive information, or inadvertent leaks \cite{CoinTelegraph2022SlopeAttack}.

\subsubsection{Shared-Custodial}
\label{sec:semi-custodial-wallets}

Shared-custodial wallets strike a balance between custodial and non-custodial models by enabling joint control of the secret key (\textcolor{teal}{\textit{sk}}) between a user and a custodian. In this setup, the \textcolor{teal}{\textit{sk}} is split or distributed across two or more parties, allowing the user to delegate a degree of transaction authorisation rights and trust to the custodian. This arrangement gives both parties partial control over the wallet's signing and recovery operations \cite{erinle2024shared, Das2024Shared-CustodialWallets}. As a result, even if one party's security is compromised, the risk of a complete \textcolor{teal}{\textit{sk}} compromise is mitigated. For example, ZenGo’s operational model implements shared custody with \acf{mpc} by storing one part of the \textcolor{teal}{\textit{sk}} on ZenGo's centralised server, while the other part remains on the user's device \cite{zengo_rec}. Other shared custodian models are discussed in \autoref{sec:design-distr}.

\subsection{Initialisation}
\label{sec:design-init}

This pertains to the creation of the wallet through \textcolor{teal}{\textit{sk}} generation (see \autoref{sec:key_generation}) or contract deployment. During initialisation in smart contract wallets, user account contracts are typically created by interactions made by the relayer. In conventional wallets, the \textcolor{teal}{\textit{sk}} generation scheme can be non-deterministic, deterministic, or hierarchical deterministic, depending on the degree of randomness and flexibility required. Another interesting design option is the \acs{kdf} choice. Typically, most wallets (e.g., Ledger \cite{ledger_seed}) employ \acf{pbkdf}; however, novel research into threshold \acf{mfkdf} construction could influence current cryptographic designs \cite{NairMulti-FactorManagement, nair2023decentralizing}. While this improves security, more processing time and power may be required to generate the derived key \cite{trezor_memory}.

\subsection{Distribution}
\label{sec:design-distr}

This is the degree of authorisation (see \autoref{sec:design-author}) or \textcolor{teal}{\textit{sk}} distribution between storage mechanisms.  Single or variations of shared authorisation between multiple user devices, multiple users or a user and a custodian (see \autoref{sec:design-cust}) are observable setups. Single setups allow for sole authorisation by a user or custodian, while authorisation is distributed in the shared setup to avoid a single point of failure. 

Multi-distributed designs typically exist in two forms: smart wallet-enabled multi-sig (on-chain multi-sig) and threshold \acs{mpc}. For smart contract wallets that follow \acf{eip} 4337, the account contract may adopt any of these schemes: single key, multi-sig, or \acs{mpc}, as the standard merely asks the contract to prove validity to \texttt{validateUserOp}. On-chain multi-sig typically has authorisation dispersed between multiple private keys (\textcolor{teal}{\textit{sk}}), while \acs{mpc} wallets divide a single \textcolor{teal}{\textit{sk}} into \enquote{key shares}, which are then distributed \cite{bip11, Lindell2020SecureComputation}. Design flexibility in some \acs{mpc} wallets also allows for a hierarchical sub-shard distribution (e.g., Web3Auth) if necessary \cite{web3_auth}. While both offer authorisation distribution, trade-offs exist between the two (see \autoref{sec:design-author} \& \autoref{sec:design-val}).

\subsection{Authentication}
\label{sec:design-authen}


Authentication is the process of verifying the legitimate wallet owner before granting access, either by decrypting \textcolor{teal}{\textit{enc\_sk}} with the \acf{kek} (see \autoref{sec:key-storage}) or by employing other methods defined within the underlying logic. Existing authentication methods include single-factor (\textcolor{teal}{\textit{pw}} or \textcolor{teal}{\textit{PIN}}), multi-factor authentication, and novel password-abstracted authentication methods such as passkeys enabled by smart contract or \acs{mpc} wallets. For instance, the Binance Web3 Wallet uses \acs{mpc} to generate three key-shares: one secured by Binance, one stored on the user’s device, and one encrypted with a user-defined recovery password and backed up to the user’s iCloud/Google Drive. The wallet uses a 2-of-3 threshold scheme to authorise transactions, so Binance’s single share is insufficient on its own \cite{Binance2023EmbracingWallet}.


\subsection{Authorisation}
\label{sec:design-author}

Authorisation in the context of wallets is defined as a direct or indirect confirmation of a state change transaction (see \hyperref[sec:def-trans]{Definition 3.2}) by a single signature or multiple signatures (\teal{$\sigma$}). In the \acs{eip}-4337 flow, the user signs a \texttt{UserOperation}. However, a Bundler/Relayer authorises the on-chain transaction by submitting the batch to the shared \texttt{EntryPoint} \cite{erc4337}. We therefore mark every 4337 wallet as \enquote{Relayer} in \autoref{tab:wlt._taxonomy}. \acs{mpc} key shards produce a single signature while being distributed among various parties with individual public addresses hidden. 

Multi-sig smart wallets demonstrate authorisation through multiple signatures, each associated with an individual public address. This approach does not enhance privacy since all involved addresses are visible on the blockchain. \acs{eip}-4337-enabled smart contract wallets employ a relayer (bundler) to aggregate multiple users' state transfer messages into a single authorised transition. Another factor that influences the authorisation setup is the choice of signature scheme.

\subsection{Validation}
\label{sec:design-val}
Transaction validation typically refers to authentication against the blockchain using the user's \textcolor{teal}{\textit{pk}} \cite{Homoliak2024SoK:Factors, Homoliak2020SmartOTPs:Wallets}. In addition to single distributed wallets, \acs{mpc} wallets also produce a single \textcolor{teal}{\textit{pk}} from key shards, which can be employed to validate the transaction. On the other hand, native multi-sig wallets validate each party's public key. \acs{eip}-4337 allows more flexible validation variations, as an \texttt{EntryPoint} contract validates and executes state changes sent by authenticated users \cite{erc4337}. Additionally, recent developments (\acs{erc}-1271 \cite{Ethereum2018ERC-1271:Contracts} \& \acs{erc}-6492 \cite{Ethereum2023ERC-6492:Contracts}) have enabled standardised and improved signature validation methods for smart contracts. 

\subsection{Recovery}
\label{sec:design-rec}

Recovery serves as a method to retrieve \textcolor{teal}{\textit{sk}} or lost transaction authorisation rights and typically follows the initialisation (see \autoref{sec:design-init}) and distribution (see \autoref{sec:design-distr}) setups selected.

\subsubsection{Seed Recovery}
The industry standard fallback for a user wallet is the \acf{bip} 39 mnemonic recovery phrase, usually 12 or 24 English words that encode the master \acf{hd} seed \cite{bip39_}. Specifically, 128–256 bits of random entropy are appended with a checksum and split into 11-bit chunks, each of which indexes one word in the 2048-word \acs{bip}-39 list. When the user re-enters the phrase, it is processed through 2048 iterations of PBKDF2 (Password-Based Key-Derivation Function v2) using HMAC-SHA-512 (a keyed-hash message-authentication code) to yield a 512-bit seed. This seed forms the root of the wallet’s \acs{hd} key tree, from which every subsequent \textcolor{teal}{\textit{sk}} is derived \cite{bip39_}.

Two notable design variations to the default mnemonics setup exist to offer additional security. First, the optional portability passphrase (\enquote{25th word}) in \acs{bip}‑39 allows plausible deniability if the base phrase is coerced \cite{vault12_bip39}. Second, \acs{slip}‑39 Shamir‑Secret‑Sharing mnemonics fragment the seed into shares, requiring a quorum (e.g., m‑of‑n) to restore the wallet \cite{trezorslip39, slip39_site}. Some mobile wallets go further by pairing mnemonics with encrypted cloud backups (e.g., Coinbase Wallet using iCloud/Google Drive), improving usability while keeping control with the user \cite{coinbase_backup}. 

Despite convenience gains, mnemonic phrases remain prime targets for social engineering and clipboard‑scraping malware, reinforcing the need for offline generation and, where feasible, distributed‑share approaches. Social platforms such as Telegram extend cloud backups into custodial-assisted models. For example, TON Space encrypts the seed locally and synchronises it with Telegram Cloud, binding recovery to the user’s Telegram ID. After re-authenticating that account, the Mini App reinjects the seed into a Wallet-V4 contract. This incurs no on-chain fee; however, it creates a single point of failure, as loss or compromise of the Telegram account threatens both availability and confidentiality \cite{beincrypto2025ton,tonwalletv4}.

\subsubsection{Seedless Recovery}
Seedless recovery eliminates mnemonic phrases and re-establishes user authorisation rights without a seed. Single or multi-party variations exist, with common instantiations including contract-based social recovery, MPC re-sharing, and other implementations such as \acf{derec} \cite{buterin_social_recovery_2021, Lindell2020SecureComputation, derec}. Implementations differ and create distinct cost profiles in smart contract and \acs{mpc} wallets. \acs{mpc} wallets perform recovery off-chain through key fragment reconstruction and thus incur no on-chain network fees. By contrast, smart contract wallets (e.g., Coinbase Smart Wallet) implement recovery as an on-chain signer/owner change that requires a network fee \cite{coinbase_smart_wallet_recovery}. However, one smart contract wallet, Argent, circumvents this by offering users off-chain recovery \cite{argent_rec}. More recently, the \acs{derec} standard proposes an interoperable, multi-party key recovery framework that allows users to regain access across different wallets and services without relying on a single custodian \cite{derec}.


\subsection{Other Design Factors}
\label{sec:design-other}

\autoref{tab:wlt._taxonomy} shows other design factors such as transparency and agnosticism. The underlying mechanism of existing hardware, software, non-custodial and shared-custodial wallets often functions in degrees of transparency. While open-source models benefit from public audits, open knowledge of mechanisms can provide an advantage to an adversary. Chain support is another important factor, as integration with multiple blockchain networks defines blockchain-agnosticism. As blockchains often operate as fragmented systems, heterogeneous designs foster enhanced interoperability.

\begin{table*}[!p]
\centering
\renewcommand{\arraystretch}{1.3}
\setlength{\tabcolsep}{1.95pt} 
\tiny
\resizebox{1.0\textwidth}{!}{
\begin{tabular}{llcccccccccccccccccccccccccccccccccccccccccccccccccccccccccccc}
\toprule
  \multicolumn{1}{c}{\textbf{Name}} &
  \multicolumn{1}{c}{\textbf{{\hyperref[fig:wallet-evolution]{Est.}}}} &
  \multicolumn{3}{c}{\textbf{{\hyperref[sec:design-cust]{Custody}}}} &
  \multicolumn{8}{c}{\textbf{{\hyperref[sec:infrastructure]{Infrastructure}}}} &
  \multicolumn{4}{c}{\textbf{{\hyperref[sec:design-init]{Initial.}}}} &
  \multicolumn{3}{c}{\textbf{{\hyperref[sec:design-distr]{Dis.}}}} & 
  \multicolumn{3}{c}{\textbf{{\hyperref[sec:design-author]{Authoris.}}}} &
  \multicolumn{3}{c}{\textbf{{\hyperref[sec:design-val]{Valid.}}}} &
  \multicolumn{5}{c}{\textbf{{\hyperref[sec:design-authen]{Authentication}}}} &
  \multicolumn{4}{c}{\textbf{{\hyperref[sec:design-rec]{Recovery}}}} &
  \multicolumn{2}{c}{\textbf{{\hyperref[sec:design-rec]{Trans.}}}} &
  \multicolumn{9}{c}{\textbf{{\hyperref[sec:design-rec]{Chains Supported}}}} &
  \multicolumn{15}{c}{\textbf{{\hyperref[sec:threat_framework]{Threat Occurrences}}}} 
  \\
\bottomrule
\end{tabular}
}
\vspace{1ex} 
\caption{
Industry wallet design variations and breadth of identified threat exposures, showing (for each wallet) the number and percentage of distinct threat categories observed (\# and \%). The denominator is the total number of threat types catalogued 15; a higher percentage means the wallet has experienced a greater diversity of threat categories. ( \smallfullcirc : include, \smallhalfcirc : part-inclusion, \smallemptycirc : not include)
}
\label{tab:wlt._taxonomy}
\end{table*}

\begin{table*}[!h]
\centering
\renewcommand{\arraystretch}{1.1}
\setlength{\tabcolsep}{1.1pt}
\footnotesize
\resizebox{\textwidth}{!}{
\begin{tabular}{%
  l@{\hspace{8pt}}
  l   
  c c 
  c c c c c c 
  l   
  c c c 
  c c   
}
\toprule
\multirow{2}{*}{\textbf{Category}}
  & \multirow{2}{*}{\textbf{Threat}}
  & \multicolumn{2}{c}{\textbf{Gap}}
  & \multicolumn{6}{c}{\textbf{Target}}
  & \multicolumn{1}{c}{\textbf{Adversary’s Capability Summary}}
  & \multicolumn{3}{c}{\textbf{Knwl.}}
  & \multicolumn{2}{c}{\textbf{Acc.}} \\
\cmidrule(lr){3-4} \cmidrule(lr){5-10} \cmidrule(lr){12-14} \cmidrule(lr){15-16}
 & 
 & \rotatebox{90}{Academia}
 & \rotatebox{90}{Incidents}
 & \rotatebox{90}{KeyGen}
 & \rotatebox{90}{TxnInit}
 & \rotatebox{90}{UserAuth}
 & \rotatebox{90}{KeyStore}
 & \rotatebox{90}{TxnSign}
 & \rotatebox{90}{TxnVer}
 & 
 & \rotatebox{90}{Public}
 & \rotatebox{90}{Restricted}
 & \rotatebox{90}{Insider}
 & \rotatebox{90}{Remote}
 & \rotatebox{90}{Physical}
\\
\midrule
\multirow{2}{*}{Network} 
  & Insecure Network Channel \cite{cve_33297, cve_14198, cve_17144}
  & {\smallfullcirc} & {\smallfullcirc}
  & {\smallemptycirc} & {\smallfullcirc} & {\smallemptycirc} & {\smallemptycirc} & {\smallemptycirc} & {\smallemptycirc}
  & Intercept transactions or deny service to wallet.
  & {\smallfullcirc} & {\smallemptycirc} & {\smallemptycirc}
  & {\smallfullcirc} & {\smallemptycirc} \\
  & Compromised Network Protocol \cite{Hu2021SecurityCountermeasures}
  & {\smallfullcirc} & {\smallemptycirc}
  & {\smallemptycirc} & {\smallemptycirc} & {\smallfullcirc} & {\smallemptycirc} & {\smallemptycirc} & {\smallemptycirc}
  & Misroute transactions via protocol flaws.
  & {\smallfullcirc} & {\smallemptycirc} & {\smallemptycirc}
  & {\smallfullcirc} & {\smallemptycirc} \\
\midrule
\multirow{7}{*}{Application}
  & Application Logic Flaw \cite{Parisi2023WalletSecurity, oren2023fireblocks}
  & {\smallfullcirc} & {\smallfullcirc}
  & {\smallemptycirc} & {\smallemptycirc} & {\smallfullcirc} & {\smallemptycirc} & {\smallemptycirc} & {\smallemptycirc}
  & Exploit the programming logic of functions.
  & {\smallfullcirc} & {\smallemptycirc} & {\smallemptycirc}
  & {\smallfullcirc} & {\smallemptycirc} \\
  & \acs{os} Vulnerabilities \cite{he2020security}
  & {\smallfullcirc} & {\smallfullcirc}
  & {\smallemptycirc} & {\smallemptycirc} & {\smallemptycirc} & {\smallfullcirc} & {\smallemptycirc} & {\smallemptycirc}
  & Exploit \acs{os} to bypass security.
  & {\smallemptycirc} & {\smallfullcirc} & {\smallemptycirc}
  & {\smallfullcirc} & {\smallemptycirc} \\
  & Library Vulnerability \cite{bitcore_lib, Ledger2023SecurityReport}
  & {\smallfullcirc} & {\smallfullcirc}
  & {\smallfullcirc} & {\smallemptycirc} & {\smallemptycirc} & {\smallemptycirc} & {\smallfullcirc} & {\smallfullcirc}
  & Exploit vulnerabilities in third-party libraries.
  & {\smallfullcirc} & {\smallfullcirc} & {\smallemptycirc}
  & {\smallfullcirc} & {\smallemptycirc} \\
  & Coding Errors \cite{Parisi2023WalletSecurity}
  & {\smallfullcirc} & {\smallfullcirc}
  & {\smallemptycirc} & {\smallfullcirc} & {\smallfullcirc} & {\smallemptycirc} & {\smallemptycirc} & {\smallemptycirc}
  & Exploit coding errors to bypass security.
  & {\smallfullcirc} & {\smallemptycirc} & {\smallemptycirc}
  & {\smallfullcirc} & {\smallemptycirc} \\
  & Insecure Interaction \cite{ZengoZengo}
  & {\smallfullcirc} & {\smallfullcirc}
  & {\smallemptycirc} & {\smallfullcirc} & {\smallemptycirc} & {\smallemptycirc} & {\smallemptycirc} & {\smallemptycirc}
  & Exploit users through UI deception.
  & {\smallfullcirc} & {\smallfullcirc} & {\smallfullcirc}
  & {\smallfullcirc} & {\smallemptycirc} \\
  & Application Provider Compromise \cite{bybit_slowmist}
  & {\smallemptycirc} & {\smallfullcirc}
  & {\smallemptycirc} & {\smallfullcirc} & {\smallemptycirc} & {\smallemptycirc} & {\smallemptycirc} & {\smallemptycirc}
  & Exploit provider infrastructure to inject code.
  & {\smallfullcirc} & {\smallemptycirc} & {\smallemptycirc}
  & {\smallfullcirc} & {\smallemptycirc} \\
  & Data Misrepresentation \cite{bybit}
  & {\smallemptycirc} & {\smallfullcirc}
  & {\smallemptycirc} & {\smallfullcirc} & {\smallemptycirc} & {\smallemptycirc} & {\smallemptycirc} & {\smallemptycirc}
  & Misrepresent transaction data on UI.
  & {\smallemptycirc} & {\smallfullcirc} & {\smallfullcirc}
  & {\smallfullcirc} & {\smallemptycirc} \\
\midrule
\multirow{2}{*}{Authentication}
  & Inadeq. Authentication \cite{Uddin2021Horus:Wallets}
  & {\smallfullcirc} & {\smallfullcirc}
  & {\smallemptycirc} & {\smallemptycirc} & {\smallfullcirc} & {\smallemptycirc} & {\smallemptycirc} & {\smallemptycirc}
  & Bypass authentication mechanisms.
  & {\smallfullcirc} & {\smallfullcirc} & {\smallemptycirc}
  & {\smallfullcirc} & {\smallfullcirc} \\
  & Low-strength Password \cite{Kiktenko2019DetectingWallets, volety2019cracking}
  & {\smallfullcirc} & {\smallfullcirc}
  & {\smallemptycirc} & {\smallemptycirc} & {\smallfullcirc} & {\smallemptycirc} & {\smallemptycirc} & {\smallemptycirc}
  & Brute-force weak passwords.
  & {\smallfullcirc} & {\smallemptycirc} & {\smallemptycirc}
  & {\smallfullcirc} & {\smallemptycirc} \\
\midrule
\multirow{8}{*}{Storage}
  & Insecure Boot Environment \cite{Shaikh2022SurveyExchanges}
  & {\smallfullcirc} & {\smallemptycirc}
  & {\smallemptycirc} & {\smallemptycirc} & {\smallemptycirc} & {\smallfullcirc} & {\smallemptycirc} & {\smallemptycirc}
  & Hijack boot to execute malicious code.
  & {\smallemptycirc} & {\smallfullcirc} & {\smallemptycirc}
  & {\smallemptycirc} & {\smallfullcirc} \\
  & Insecure Permissions \cite{cve_32969, halborn_vuln}
  & {\smallemptycirc} & {\smallfullcirc}
  & {\smallemptycirc} & {\smallemptycirc} & {\smallemptycirc} & {\smallfullcirc} & {\smallemptycirc} & {\smallemptycirc}
  & Escalate file-system permissions.
  & {\smallemptycirc} & {\smallfullcirc} & {\smallemptycirc}
  & {\smallfullcirc} & {\smallfullcirc} \\
  & Inadequate Encryption \cite{cve_15947, CoinTelegraph2022SlopeAttack}
  & {\smallfullcirc} & {\smallfullcirc}
  & {\smallfullcirc} & {\smallemptycirc} & {\smallemptycirc} & {\smallfullcirc} & {\smallfullcirc} & {\smallemptycirc}
  & Read unencrypted secrets at rest.
  & {\smallemptycirc} & {\smallfullcirc} & {\smallfullcirc}
  & {\smallfullcirc} & {\smallfullcirc} \\
  & Data Remanence \cite{trezor_memory, trezor_medium}
  & {\smallfullcirc} & {\smallfullcirc}
  & {\smallemptycirc} & {\smallemptycirc} & {\smallemptycirc} & {\smallfullcirc} & {\smallemptycirc} & {\smallemptycirc}
  & Recover keys from memory remnants.
  & {\smallemptycirc} & {\smallfullcirc} & {\smallemptycirc}
  & {\smallemptycirc} & {\smallfullcirc} \\
  & Data Manipulation \cite{trezor_memory, trezor_medium}
  & {\smallfullcirc} & {\smallfullcirc}
  & {\smallemptycirc} & {\smallemptycirc} & {\smallemptycirc} & {\smallfullcirc} & {\smallemptycirc} & {\smallemptycirc}
  & Tamper with stored data.
  & {\smallemptycirc} & {\smallfullcirc} & {\smallemptycirc}
  & {\smallfullcirc} & {\smallfullcirc} \\
  & Micro-electrical Exposure \cite{courbon2016reverse}
  & {\smallfullcirc} & {\smallfullcirc}
  & {\smallemptycirc} & {\smallemptycirc} & {\smallemptycirc} & {\smallfullcirc} & {\smallemptycirc} & {\smallemptycirc}
  & Probe chip side-effects.
  & {\smallemptycirc} & {\smallfullcirc} & {\smallemptycirc}
  & {\smallemptycirc} & {\smallfullcirc} \\
  & Storage Provider Compromise \cite{CoinTelegraph2022SlopeAttack}
  & {\smallemptycirc} & {\smallfullcirc}
  & {\smallemptycirc} & {\smallemptycirc} & {\smallemptycirc} & {\smallfullcirc} & {\smallemptycirc} & {\smallemptycirc}
  & Breach external storage vendor.
  & {\smallemptycirc} & {\smallfullcirc} & {\smallfullcirc}
  & {\smallfullcirc} & {\smallemptycirc} \\
\midrule
\multirow{3}{*}{Cryptanalysis}
  & Predictable \acs{rng} \cite{cve_31290, cve_23660}
  & {\smallfullcirc} & {\smallfullcirc}
  & {\smallfullcirc} & {\smallemptycirc} & {\smallemptycirc} & {\smallemptycirc} & {\smallemptycirc} & {\smallemptycirc}
  & Predict or replay RNG outputs.
  & {\smallfullcirc} & {\smallfullcirc} & {\smallemptycirc}
  & {\smallfullcirc} & {\smallemptycirc} \\
  & Weak Signature \cite{Rokhjavan2023SecuringWallets}
  & {\smallfullcirc} & {\smallfullcirc}
  & {\smallemptycirc} & {\smallemptycirc} & {\smallemptycirc} & {\smallemptycirc} & {\smallfullcirc} & {\smallfullcirc}
  & Forge signatures under weak crypto.
  & {\smallfullcirc} & {\smallfullcirc} & {\smallemptycirc}
  & {\smallfullcirc} & {\smallemptycirc} \\
  & Side-channel Leakage \cite{cve_14353, cve_14354, KrakenBlog}
  & {\smallfullcirc} & {\smallfullcirc}
  & {\smallemptycirc} & {\smallemptycirc} & {\smallemptycirc} & {\smallfullcirc} & {\smallemptycirc} & {\smallemptycirc}
  & Extract secrets via side-channels.
  & {\smallfullcirc} & {\smallfullcirc} & {\smallemptycirc}
  & {\smallfullcirc} & {\smallfullcirc} \\
\midrule
\multirow{2}{*}{Other}
  & Insider Collusion \cite{decrypt_ftx}
  & {\smallemptycirc} & {\smallfullcirc}
  & {\smallemptycirc} & {\smallemptycirc} & {\smallfullcirc} & {\smallfullcirc} & {\smallemptycirc} & {\smallemptycirc}
  & Collude with insiders.
  & {\smallemptycirc} & {\smallemptycirc} & {\smallfullcirc}
  & {\smallfullcirc} & {\smallfullcirc} \\
  & Insider Compromise \cite{Ledger2023SecurityReport}
  & {\smallemptycirc} & {\smallfullcirc}
  & {\smallemptycirc} & {\smallemptycirc} & {\smallfullcirc} & {\smallfullcirc} & {\smallemptycirc} & {\smallemptycirc}
  & Abuse insider access.
  & {\smallemptycirc} & {\smallemptycirc} & {\smallfullcirc}
  & {\smallfullcirc} & {\smallfullcirc} \\
\bottomrule
\end{tabular}
}
\normalsize
\vspace{1ex}
\caption{Classification of threats (\autoref{sec:threat_class}) showing the targetted operation, and the capabilities (\autoref{sec:adversary_cap}), knowledge and accessibility of the adversary. A gap analysis of threats is also conducted to compare industry and academia.}
\label{tab:threat_capability}
\end{table*}



\section{Threat Model}
\label{sec:threat_framework}

We analyse threats to the wallet mechanism, considering adversary goals, knowledge, and capabilities. Using our design taxonomy (\autoref{tab:wlt._taxonomy}), we also identify industry threats and highlight gaps between industry and academia (\autoref{tab:threat_capability}).

\subsection{Classification}
\label{sec:threat_class}

Our threat classification is structured around distinct operations within the wallet mechanism across three stages: wallet initialisation, transaction generation, and transaction broadcast. Threats to the system can be categorised into five areas: network, authentication, application, storage and memory, and cryptanalysis.

\subsubsection{Network}
\label{sec:vuln_mech}

The wallet communicates with the blockchain to retrieve and broadcast \textcolor{teal}{\textit{state\_trans\_info}} using internet protocols. The network enables the secure transmission of messages within and outside of the system. Vulnerabilities in the communication channels can be targeted, as shown in \autoref{tab:attack_vectors}. Service providers in the network can also be compromised, rendering messages vulnerable to interception and alteration.

\subsubsection{Application}
\label{sec:vuln_mech}

Wallets rely on application libraries \cite{Hu2021SecurityCountermeasures}, and operating systems \cite{he2020security, li2020android}, which may possess vulnerabilities that the adversary can exploit. Vulnerabilities in these systems include application logic vulnerabilities such as key recovery \cite{cve_15302}, signature verification \cite{cve_14199}, and input validation \cite{immunefi} flaws, which can result in privilege escalation. Additionally, malware exposure \cite{balakrishnan2023analysis, li2020android}, insecure third-party interactions \cite{ZengoZengo, thodex}, and user negligence \cite{weichbroth2023security} can threaten the security of the \textcolor{teal}{\textit{sk}}, \textcolor{teal}{\textit{rdm\_seed}}, or \textcolor{teal}{\textit{pw}}. For instance, projects integrating TON wallets have experienced silent exfiltration of mnemonic phrases via malicious application libraries \cite{sockettonpkg, ghsadvisory}. Web3 wallets embedded in social platforms amplify supply-chain risk. Malicious npm modules impersonating TON SDKs (e.g., \texttt{@ton-wallet/create}) execute clipboard-sniffers that forward seed phrases to attacker-controlled Telegram bots \cite{sockettonpkg,beincrypto2024socket}. Since wallet logic is tightly coupled to chatbot \acs{api}s, a single rogue Mini App link can invoke in-chat transaction authorisation by users, as seen in the June 2025 phishing wave \cite{slowmisttron}.

\begin{figure}[!t]
	\centering \includegraphics[scale=0.70]
 {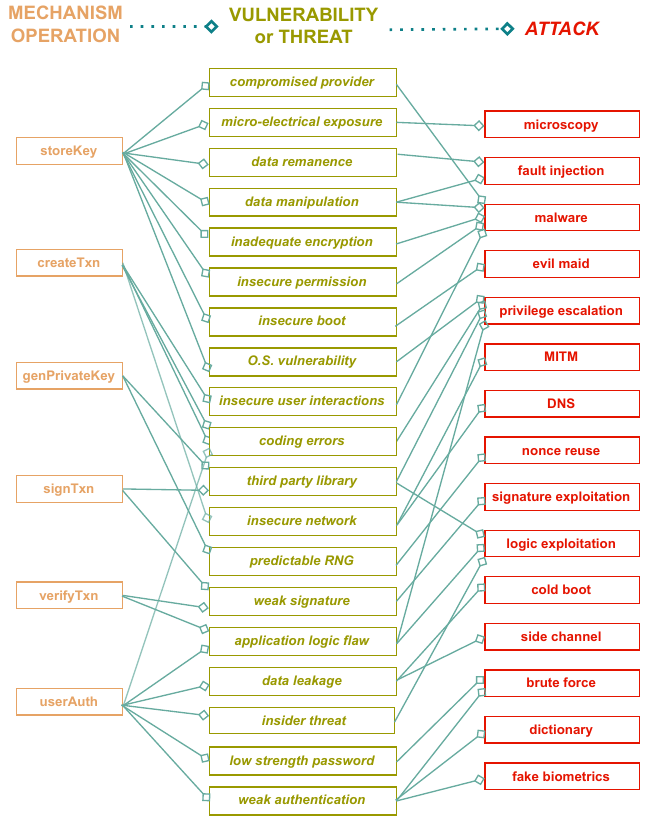}
	\caption{Mapping of the wallet mechanism (\autoref{sec:wallet_mechanism}) to threats/vulnerability occurrences (\autoref{sec:threat_framework}) and attack methods (\autoref{sec:attack-framework}).}
	\label{fig:wallet-mapping}
\end{figure}

\subsubsection{Authentication}
\label{sec:vuln_mech}

Authentication is a critical process in modern wallets, as only an authorised owner can decrypt an encrypted private key (\textcolor{teal}{\textit{enc\_sk}}) and sign transactions (refer to the \textcolor{orange}{\textit{encrypt}} and \textcolor{orange}{\textit{decrypt}} functions in \hyperref[algo:cryptocurrency-wallet]{Algorithm 1} and \hyperref[algo:transaction-signing]{Algorithm 2}, respectively). Authentication attacks aim to compromise the wallet function that verifies the user's identity, thereby gaining unauthorised access to wallets. The authentication functions, which handle the encryption and decryption of the \textcolor{teal}{\textit{enc\_sk}}, can be vulnerable to insecure boot environments \cite{Shaikh2022SurveyExchanges} and single-factor authentication methods and low-strength passwords (\textcolor{teal}{\textit{pw}}).

\subsubsection{Storage and Memory}
\label{sec:vuln_phy}

Data stored can be vulnerable to threats of extraction, manipulation and disruption. Exploitation of the wallet's storage mechanism (see \autoref{sec:key-storage}) can lead to the compromise of \textcolor{teal}{\textit{sk}}, \textcolor{teal}{\textit{rdm\_seed}} or \textcolor{teal}{\textit{pw}}. Storage mechanism vulnerabilities include data remanence \cite{Shaikh2022SurveyExchanges}, unencrypted data \cite{breier2022practical, robinson2022new}, and physical security vulnerabilities \cite{courbon2016reverse} that can be exploited by the adversary.

\subsubsection{Cryptanalysis}
\label{sec:vuln_mech}

Cryptographic vulnerabilities may exist in the signature scheme (\textcolor{orange}{\textit{keyGen}}, \textcolor{orange}{\textit{txnSign}}, \textcolor{orange}{\textit{txnVer}}) as a result of the direct implementation or unintended data leakages from side channels. These vulnerabilities include hash function vulnerabilities \cite{shrivas2020disruptive}, weak signatures (\teal{$\sigma$}) \cite{Rokhjavan2023SecuringWallets}, predictable \acf{rng} \cite{brengel2018identifying}, and data leakages from side-channels \cite{Park2023, kocher1996timing}.

\subsubsection{Other Threats}
\label{sec:vuln_ext}

Threats can occur via other avenues, such as an insider who may have access to transactional information, user credentials and other security details. These can arise from insiders acting maliciously or by exploitation through coercion or social engineering methods. Custodial (\autoref{sec:custodial-wallets}) and Shared-custodial (\autoref{sec:semi-custodial-wallets}) architectures are more vulnerable to these threats due to their more centralised architecture. Non-custodial setups (see \autoref{sec:non-custodial-wallets}) may be vulnerable if third-party services are employed for functionalities such as \textcolor{teal}{\textit{pw}} management or if inadequate access controls are relied upon (e.g., Ledger incident \cite{zerocap}).

\subsection{Adversary's Goals}
\label{sec:adversary_goal}

 We define an adversary, \textcolor{teal}{\textit{A}}, who aims to exploit threats described above to trigger unauthorised transactions to an adversary-controlled wallet address or disrupt operations. The major goals of \textcolor{teal}{\textit{A}} include:
\begin{itemize}
    \item \textbf{Credential Compromise:} \textcolor{teal}{\textit{A}} aims to compromise \textcolor{teal}{\textit{sk}}, \textcolor{teal}{\textit{rdm\_seed}} and \textcolor{teal}{\textit{pw}} by exploiting wallet mechanism vulnerabilities or user-interactions.
    \item \textbf{State Transition Information Manipulation:} \textcolor{teal}{\textit{A}} aims to modify the \textcolor{teal}{\textit{state\_trans\_info}} created by the user such as \textcolor{teal}{\textit{recipient\_address}}. Following this, \textcolor{teal}{\textit{A}} deceives the user into signing the transaction. \textcolor{teal}{\textit{A}} may also manipulate the \textcolor{teal}{\textit{state\_trans\_info}} displayed on the wallet interface
\end{itemize}

\subsection{Adversary's Capabilities}
\label{sec:adversary_cap}

\autoref{tab:threat_capability} details the various capabilities of \textcolor{teal}{\textit{A}}, illustrating how identified vulnerabilities can be exploited to achieve an objective with various degrees of knowledge and access. \textcolor{teal}{\textit{A}} can possess public, restricted and insider knowledge. Public knowledge includes information that is openly accessible to anyone, such as open-source code, publicly available audit reports, discussions in open forums, websites, and applications. Restricted knowledge refers to information that is not readily accessible to the public and often requires specific roles, permissions, or effort to obtain. Information that is only accessible to individuals within an organisation is defined as insider knowledge, particularly in setups where custodians have some level of authorisation (\autoref{sec:design-cust}). \textcolor{teal}{\textit{A}} can also execute several attack capabilities remotely or physically.

\section{Attack Taxonomy}
\label{sec:attack-framework}

\begin{figure}[!t]
	\centering \includegraphics[scale=0.85]
 {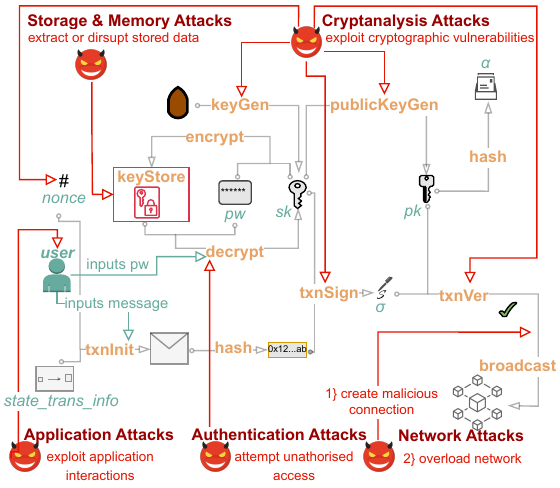}
	\caption{Attack classification on wallet mechanism showing targeted operations and components (see \autoref{tab:attack_vectors}).}
	\label{fig:wallet-attacks}
\end{figure}

In this section, we present a comprehensive taxonomy of wallet attack vectors, systematically examining the methods, techniques, and targeted components involved. Building on our generalised wallet mechanisms and threat model taxonomy, we outline a broad spectrum of attacks, as illustrated in \autoref{fig:wallet-attacks}. These attacks are categorised according to the specific functions and components targeted within the wallet infrastructure (see \autoref{sec:wallet_mechanism}) and the threats exploited (see \autoref{sec:threat_class}). We further incorporate the infrastructure layer of our design taxonomy to capture the multi-layered nature of these threats, as summarised in \autoref{tab:attack_vectors}.

\subsection{Network}
\label{sec:network-attacks} 

\subsubsection{Connection Hijack}
\label{sec:mitm}

These attacks aim to compromise the communication channel between wallets and other network participants using \acf{mitm} attacks to intercept and modify the \textcolor{teal}{\textit{txn}} message generated by \hyperref[algo:transaction-signing]{Algorithm 2}. Various types of \acs{mitm} attacks include Rogue \acs{ap} \cite{Hu2021SecurityCountermeasures}, \acs{dns} spoofing \cite{Ahmed2017MitigatingNetworking, Al-Mashhadi2020ASystems}, \acs{ip} spoofing \cite{shrivas2020disruptive}, and \acf{bgp} hijacking \cite{ekparinya2018impact}, as shown in \autoref{tab:attack_vectors}. Hardware wallets are vulnerable to these attacks if the online wallet client (see \autoref{sec:hardware-wallets}) is compromised. Ledger has previously reported susceptibility to \acs{mitm} attacks.

The Rogue \acf{ap} vector functions through unauthorised WiFi hotspots that can intercept transactions by exploiting the \textcolor{orange}{\textit{txnInit}} function. This allows an attacker to modify \textcolor{teal}{\textit{state\_trans\_info}} before blockchain forwarding, potentially redirecting funds to a different address than the recipient's address \cite{Hu2021SecurityCountermeasures}. The \acf{dns} spoofing vector occurs when a \acs{dns} resolver, which translates human-readable domain names into IP addresses, is compromised \cite{pillai2019smart}. This leads to fraudulent cryptocurrency service website redirection. One notable example is the 2017 EtherDelta DNS hijack, where attackers altered \acs{dns} records to redirect users to a phishing clone \cite{CryptocurrencyScheme}. An attacker can also execute a \acf{bgp} hijacking attack that maliciously advertises false \acs{bgp} routes to divert traffic intended for legitimate blockchain nodes (see \hyperref[algo:transaction-broadcast]{Algorithm 3}) or wallet \acs{api} endpoints \cite{ekparinya2018impact}. The MyEtherWallet attacker employed the \acs{bgp} hijacking vector \cite{myetherwallet}. Another connection hijack avenue, the \acf{arp} spoofing vector, is initiated when attackers broadcast fraudulent \acs{arp} messages across a local network. This links their MAC address with the \acs{ip} address of a legitimate network host to redirect the user's transaction data generated in \hyperref[algo:transaction-signing]{Algorithm 2} \cite{Hu2021SecurityCountermeasures, ekparinya2018impact}.


\subsubsection{Service Denial}
\label{sec:dos}

This is executed using adversary-controlled devices to orchestrate \acf{ddos} attacks which overwhelm the network infrastructure with an excessive volume of requests, causing a decline or cessation of wallet operations (see \autoref{sec:wallet_mechanism}) \cite{ChandanProtectiveCryptocoin}. These attacks often target the \acf{icmp}, \acf{tcp} handshake mechanism, and other network infrastructure \cite{chaganti2022comprehensive}. One common medium of conducting a \acs{ddos} attack is through botnets, which involves an adversary using a network of computers \cite{krombholz2015advanced}.

The \acf{icmp} flooding vector overloads a wallet with network requests (\acs{icmp} echo request packets) at a rate exceeding the processing capacity. This results in a decline or cessation of transaction management operations (see \autoref{sec:transaction_management}) \cite{chaganti2022comprehensive}. An adversary may also disrupt the wallet network by exploiting the \acf{tcp} handshake mechanism, which establishes a connection between the wallet application and its servers, through \acf{syn} attacks \cite{chaganti2022comprehensive}.

\subsection{Application}
\label{sec:application-attacks}

\subsubsection{Malware Execution}
\label{sec:malware}

This attack intrusively exploits system vulnerabilities to steal transaction data, the \textcolor{teal}{\textit{sk}} and password credentials, or to manipulate wallet operations as described in \autoref{sec:wallet_mechanism}. Malware threatens the wallet mechanism by replacing the \textcolor{teal}{\textit{recipient\_address}} via clipboard hijackers \cite{li2020android} or through input monitoring via keyloggers \cite{balakrishnan2023analysis} and other spyware types \cite{weichbroth2023security, ferdous2023review}. Hardware wallets are also vulnerable to clipboard hijack attacks \cite{ivanov2021ethclipper, Akter2023AChallenges}; malware can be injected through interactions between the wallet and removable media such as USB drives \cite{guri2018beatcoin}. 

Malware can also be engineered to monitor user actions and retrieve the user's password (\textcolor{teal}{\textit{pw}}) or private key (\textcolor{teal}{\textit{sk}}) \cite{weichbroth2023security, ferdous2023review}. Spyware includes keyloggers which can track every keystroke executed on an infected wallet device to steal confidential data \cite{Shaikh2022SurveyExchanges, balakrishnan2023analysis}. The custodial wallet Cashaa \cite{CoinTelegraph} and non-custodial wallets BitKeep \cite{CertiKIncidents} and Bittensor \cite{Explained:2024} have previously been exploited by malware-based vectors. Malware can also be combined with other attack methods, such as social engineering or privilege escalation, to achieve hacks, as noted in the ByBit case (see \autoref{sec:bybit_case} and \autoref{fig:lossMonthly}).

\begin{figure*}[t]
    \centering
    \includegraphics[width=2.15\columnwidth]{Images/Timeline_of_Wallet_Hacks.pdf}
    \caption{Notable wallet incidents (in million USD) between 2012-01 and 2025-04, classified on the custody axis (\autoref{sec:custodial-wallets}). More detail is provided in \autoref{tab:attack-incidents}.}
    \label{fig:lossMonthly}
\end{figure*}

\subsubsection{Social Engineering}
\label{sec:social}
These attacks aim to manipulate the user to divulge confidential data. Phishing attacks, for instance, aim to deceive wallet users into revealing \textcolor{teal}{\textit{sk}} or \textcolor{teal}{\textit{pw}} by mimicking legitimate services. Once successful, attackers can leverage additional vectors to gain unauthorised access \cite{krombholz2015advanced}.

Notably, malware delivered through phishing, such as Pink Drainer, Monkey Drainer, Venom Drainer, and Inferno, has been particularly effective against non-custodial wallets (see \autoref{tab:attack-incidents}). Phishing attacks have also been effective against custodial wallets \cite{HTXReport, HackScience} and notable individuals \cite{Explained:2024}. Adversaries have also exploited third-party dependencies by targeting their personnel, thereby extending the reach of social engineering campaigns \cite{bybit}. 

Telegram-embedded wallets heighten social-engineering exposure. Coordinated Telegram bots and rogue Mini App have drained millions from users \cite{slowmisttron,dailycoin}. Address-poisoning adds yet another twist: attackers inject look-alike addresses into a victim’s history so that a routine copy-and-paste transaction quietly redirects funds \cite{MetaMaskScam}.

\subsubsection{Privilege Escalation}
\label{sec:privilege}

These attacks aim to circumvent standard access controls to acquire elevated permissions. In Android root privilege attacks, the adversary can gain unauthorised root access to mobile wallets through vulnerabilities in the \acf{os} \cite{he2020security}. Another OS-related attack, Android USB debugging \cite{he2020security}, exploits \acf{os} vulnerabilities in mobile devices by wireless debugging, using a computer connected to the same network. Following this, the adversary gains unrestricted access to manipulate the execution flow of the wallet and capture \textcolor{teal}{\textit{sk}}, \textcolor{teal}{\textit{rdm\_seed}}, and other sensitive data \cite{he2020security}.

\begin{table}[!htbp]
\centering
\tiny
\setlength{\tabcolsep}{2.1pt}
\renewcommand{\arraystretch}{0.9}
\begin{tabular}{lllrll}
\toprule
\textbf{Name} & \textbf{Custody Design} & \textbf{Date} & \textbf{Loss (\$)} & \textbf{Attack Category} & \textbf{Attack Name} \\
\midrule
ByBit \cite{bybit}  
  & Custodial  
  & 
    2025-02 
  & 1,500M  
  & Application  
  & Logic Exploitation \\

US Govt. \cite{Decrypt}  
  & Non-Custodial  
  & 
    2024-10 
  & 50M  
  & –  
  & – \\

BigX \cite{Explained:2024}  
  & Custodial  
  & 
    2024-09 
  & 52M  
  & –  
  & – \\

Indodax \cite{IndonesianTRX}  
  & Custodial  
  & 
    2024-09 
  & 22M  
  & –  
  & – \\

WazirX \cite{Explained:2024g}  
  & Custodial  
  & 
    2024-07 
  & 235M  
  & Application  
  & Logic Exploitation \\

Bittensor \cite{Explained:2024}  
  & Non-Custodial  
  & 
    2024-07 
  & 8M  
  & Application  
  & Malware \\

BTCTurk \cite{Explained:2024}  
  & Custodial  
  & 
    2024-06 
  & 55M  
  & –  
  & – \\

Loopring \cite{Explained:2024}  
  & Non-Custodial  
  & 
    2024-06 
  & 5M  
  & Authentication  
  & Identity Spoofing\textsuperscript{*} \\

Lykke \cite{CoinTelegraph}  
  & Custodial  
  & 
    2024-06 
  & 22M  
  & –  
  & – \\

DMM Bitcoin \cite{Explained:2024}  
  & Custodial  
  & 
    2024-05 
  & 305M  
  & –  
  & – \\

Axie Co-Founder \cite{Decrypt}  
  & Non-Custodial  
  & 
    2024-02 
  & 10M  
  & –  
  & – \\

Fixed Float \cite{Explained:2024}  
  & Custodial  
  & 
    2024-02 
  & 26.1M  
  & –  
  & – \\

kirilm.eth \cite{Explained:2024}  
  & Non-Custodial  
  & 
    2024-02 
  & 5.1M  
  & Application  
  & Phishing \\

Ripple Co-Founder \cite{RippleMillion}  
  & Non-Custodial  
  & 
    2024-01 
  & 112.5M  
  & –  
  & – \\

HTX (Huobi) \cite{HTXReport}  
  & Custodial  
  & 
    2023-11 
  & 13.6M  
  & –  
  & \teal{\textit{sk}} Compromise\textsuperscript{*} \\

Pink Drainer \cite{RektREKT}  
  & Non-Custodial  
  & 
    2023-11 
  & 12M  
  & Application  
  & Phishing, Malware \\

Monkey Drainer \cite{RektREKT}  
  & Non-Custodial  
  & 
    2023-11 
  & 16M  
  & Application  
  & Phishing, Malware \\

Venom Drainer \cite{RektREKT}  
  & Non-Custodial  
  & 
    2023-11 
  & 27M  
  & Application  
  & Phishing, Malware \\

Infarno \cite{infarno}  
  & Non-Custodial  
  & 
    2023-11 
  & 66M  
  & Application  
  & Phishing, Malware \\

Poloniex \cite{RektREKT}  
  & Custodial  
  & 
    2023-11 
  & 126M  
  & –  
  & \teal{\textit{sk}} Compromise\textsuperscript{*} \\

Lastpass \cite{RektREKT}  
  & Non-Custodial  
  & 
    2023-10 
  & 37M  
  & Authentication  
  & – \\

Fantom Fdn. \cite{AnalysisMedium}  
  & Non-Custodial  
  & 
    2023-10 
  & 7M  
  & –  
  & – \\

HTX (Huobi) \cite{HTXReport}  
  & Custodial  
  & 
    2023-09 
  & 8M  
  & Application  
  & Phishing \\

Fake Voucher \cite{RektREKT}  
  & Non-Custodial  
  & 
    2023-09 
  & 4.5M  
  & Application  
  & Phishing \\

Remitano \cite{RektREKT}  
  & Custodial  
  & 
    2023-09 
  & 2.7M  
  & Application  
  & – \\

CoinEx \cite{CoinTelegraph}  
  & Custodial  
  & 
    2023-09 
  & 55M  
  & –  
  & \teal{\textit{sk}} Compromise\textsuperscript{*} \\

Monero \cite{MoneroFlash}  
  & Non-Custodial  
  & 
    2023-09 
  & 0.5M  
  & –  
  & – \\

AlphaPo \cite{RektREKT}  
  & Custodial  
  & 
    2023-07 
  & 60M  
  & –  
  & \teal{\textit{sk}} Compromise\textsuperscript{*} \\

Atomic Wallet \cite{CoinTelegraph}  
  & Non-Custodial  
  & 
    2023-06 
  & 100M  
  & –  
  & – \\

Bitrue \cite{Explained:2024}  
  & Custodial  
  & 
    2023-04 
  & 23M  
  & –  
  & \teal{\textit{sk}} Compromise\textsuperscript{*} \\

GDAC \cite{CoinTelegraph}  
  & Custodial  
  & 
    2023-04 
  & 13M  
  & –  
  & \teal{\textit{sk}} Compromise\textsuperscript{*} \\

MyAlgo \cite{CoinTelegraph}  
  & Non-Custodial  
  & 
    2023-02 
  & 9.2M  
  & –  
  & – \\

BitKeep \cite{CertiKIncidents}  
  & Non-Custodial  
  & 
    2022-12 
  & 8M  
  & Application  
  & Phishing, Malware \\

FTX \cite{FTXMistake}  
  & Custodial  
  & 
    2022-11 
  & 450M  
  & Authentication  
  & Sim Swap Attack \\

Deribit \cite{CryptoWithdrawals}  
  & Custodial  
  & 
    2022-11 
  & 28M  
  & Application  
  & – \\

Wintermute \cite{TheMedium}  
  & Custodial  
  & 
    2022-09 
  & 160M  
  & Authentication  
  & Brute force \\

Slope \cite{CoinTelegraph}  
  & Non-Custodial  
  & 
    2022-08 
  & 8M  
  & Storage and Memory  
  & – \\

MetaMask \cite{CertiKIncidents}  
  & Non-Custodial  
  & 
    2022-04 
  & 0.65M  
  & Authentication  
  & Phishing \\

Crypto.com \cite{Explained:2024}  
  & Custodial  
  & 
    2022-01 
  & 30M  
  & Authentication  
  & – \\

Lympo \cite{CoinTelegraph}  
  & Custodial  
  & 
    2022-01 
  & 18.7M  
  & –  
  & – \\

LCX \cite{LookingHacken}  
  & Custodial  
  & 
    2022-01 
  & 8M  
  & –  
  & \teal{\textit{sk}} Compromise\textsuperscript{*} \\

Vulcan Forged \cite{VulcanHack}  
  & Non-Custodial  
  & 
    2021-12 
  & 140M  
  & Application  
  & \teal{\textit{sk}} Compromise\textsuperscript{*} \\

BitMart \cite{HackScience}  
  & Custodial  
  & 
    2021-12 
  & 196M  
  & Application  
  & Phishing \\

Liquid \cite{HackBreach}  
  & Custodial  
  & 
    2021-08 
  & 90M  
  & Application  
  & \teal{\textit{sk}} Compromise\textsuperscript{*} \\

Roll \cite{CoinDesk}  
  & Custodial  
  & 
    2021-03 
  & 5.7M  
  & Application  
  & \teal{\textit{sk}} Compromise\textsuperscript{*} \\

MetaMask \cite{Explained:2024}  
  & Non-Custodial  
  & 
    2020-12 
  & 8M  
  & –  
  & – \\

KuCoin \cite{kucoinNew}  
  & Custodial  
  & 
    2020-09 
  & 275M  
  & Application  
  & \teal{\textit{sk}} Compromise\textsuperscript{*} \\

Cashaa \cite{CoinTelegraph}  
  & Custodial  
  & 
    2020-07 
  & 3.1M  
  & Application  
  & Malware \\

Trinity Wallet \cite{IOTA:Wallet}  
  & Non-Custodial  
  & 
    2020-02 
  & 2.3M  
  & Application  
  & – \\

Altsbit \cite{AltsbitZDNET}  
  & Custodial  
  & 
    2020-02 
  & 72.5M  
  & Application  
  & – \\

Upbit \cite{UpbitMedium}  
  & Custodial  
  & 
    2019-11 
  & 49M  
  & Application  
  & Phishing, Malware \\

Bitpoint \cite{BitPointMedium}  
  & Custodial  
  & 
    2019-07 
  & 36.5M  
  & –  
  & – \\

Vindax \cite{VinDAXBlock}  
  & Custodial  
  & 
    2019-11 
  & 0.5M  
  & –  
  & – \\

Bitrue \cite{CryptoNews}  
  & Custodial  
  & 
    2019-06 
  & 4.5M  
  & Authentication  
  & – \\

Gatehub \cite{OverviewMedium}  
  & Custodial  
  & 
    2019-06 
  & 9.5M  
  & –  
  & – \\

Binance Exchange \cite{binanceNew}  
  & Custodial  
  & 
    2019-05 
  & 40M  
  & Unknown  
  & – \\

Bithumb \cite{CoinDesk}  
  & Custodial  
  & 
    2019-03 
  & 13M  
  & Other  
  & Insider Job \\

Coinbene \cite{CoinTelegraph}  
  & Custodial  
  & 
    2019-03 
  & 99M  
  & –  
  & – \\

DragonEX \cite{CoinDesk}  
  & Custodial  
  & 
    2019-03 
  & 1M  
  & Application  
  & – \\

Cryptopia \cite{HowHacken}  
  & Custodial  
  & 
    2019-02 
  & 16M  
  & –  
  & \teal{\textit{sk}} Compromise\textsuperscript{*} \\

LocalBitcoins \cite{CoinDesk}  
  & Custodial  
  & 
    2019-01 
  & 0.02M  
  & Application  
  & Phishing \\

Electrum \cite{DeepSwig}  
  & Non-Custodial  
  & 
    2018-12 
  & 0.75M  
  & Application  
  & Phishing \\

Maplechange \cite{MapleChangeInvestorPlace}  
  & Custodial  
  & 
    2018-10 
  & 6M  
  & –  
  & – \\

Zaif \cite{CoinDesk}  
  & Custodial  
  & 
    2018-09 
  & 100M  
  & –  
  & – \\

Coinrail \cite{CoinDesk}  
  & Custodial  
  & 
    2018-06 
  & 40M  
  & –  
  & – \\

MyEtherWallet \cite{myetherwallet}  
  & Non-Custodial  
  & 
    2018-04 
  & 0.15M  
  & Network  
  & \acs{bgp} Hijacking \\

Gate.io \cite{ZachXBTWraps}  
  & Custodial  
  & 
    2018-04 
  & 234M  
  & –  
  & – \\

CoinSecure \cite{CoinDesk}  
  & Custodial  
  & 
    2018-04 
  & 3.5M  
  & Other  
  & Insider Job \\

Bitgrail \cite{BitGrailCoinMarketCap}  
  & Custodial  
  & 
    2018-02 
  & 146M  
  & Other  
  & Insider Job \\

CoinCheck \cite{TheHack}  
  & Custodial  
  & 
    2018-01 
  & 560M  
  & –  
  & – \\

BlackWallet \cite{BlackWalletFault}  
  & Non-Custodial  
  & 
    2018-01 
  & 0.4M  
  & Network  
  & \acs{dns} Spoofing \\

EtherDelta \cite{CryptocurrencyScheme}  
  & Custodial  
  & 
    2017-12 
  & 1.4M  
  & Network  
  & \acs{dns} Spoofing \\

Parity \cite{palladino2017parity}  
  & Non-Custodial  
  & 
    2017-07 
  & 30M  
  & Application  
  & Logic Exploitation \\

Yapizon \cite{CoinTelegraph}  
  & Custodial  
  & 
    2017-04 
  & 5.3M  
  & –  
  & – \\

Bitfinex \cite{CoinDesk}  
  & Custodial  
  & 
    2016-08 
  & 623M  
  & Application  
  & – \\

Gatecoin \cite{CoinDesk}  
  & Custodial  
  & 
    2016-05 
  & 2.1M  
  & –  
  & – \\

Shapeshift \cite{LootingShapeShift}  
  & Custodial  
  & 
    2016-04 
  & 0.23M  
  & Other  
  & Insider Job \\

Bitstamp \cite{DetailsRevealed}  
  & Custodial  
  & 
    2015-12 
  & 5M  
  & Application  
  & Phishing \\

BTER \cite{CoinDesk}  
  & Custodial  
  & 
    2015-08 
  & 1.65M  
  & Application  
  & – \\

Mintpal \cite{RememberingLedger}  
  & Custodial  
  & 
    2014-07 
  & 2M  
  & Other  
  & Insider Job \\

Poloniex \cite{PoloniexHack}  
  & Custodial  
  & 
    2014-03 
  & 0.05M  
  & Application  
  & – \\

Mt. Gox \cite{mtgox_hack}  
  & Custodial  
  & 
    2014-02 
  & 460M  
  & –  
  & – \\

Bitcash \cite{CzechEmptied}  
  & Custodial  
  & 
    2013-11 
  & 0.1M  
  & Application  
  & Phishing \\

Bitfloor \cite{HackSecurityWeek}  
  & Custodial  
  & 
    2012-09 
  & 0.25M  
  & Application  
  & \teal{\textit{sk}} Compromise\textsuperscript{*} \\

Bitcoinica \cite{ExchangeStolen}  
  & Custodial  
  & 
    2012-03 
  & 0.09M  
  & Application  
  & \teal{\textit{sk}} Compromise\textsuperscript{*} \\

\midrule
\textbf{Summary:}
  & \textbf{85 incidents}
  & \textbf{2012–2025}
  & \textbf{6.98B}
  &  
  &  
\\
\bottomrule
\end{tabular}
\caption{Wallet attack incidents in the industry. We retrieve 85 notable attack incidents involving both custodial and non-custodial wallets. Several attack methods remain unknown (–) or undetailed, we indicate undetailed incidents with \textsuperscript{*}.}
\label{tab:attack-incidents}
\end{table}

\subsubsection{Logic Exploitation}
\label{sec:logic_expl}

Logic flow exploitation encompasses several wallet types and involves identifying and exploiting flaws in the programming logic of a wallet mechanism (\autoref{sec:wallet_mechanism}) to gain unauthorised access or manipulate wallet functions \cite{Parisi2023WalletSecurity}. Notable incidents include WazirX (2024), where investigators linked the drain to a malicious Safe module that slipped through the upgrade mechanism and rewired the wallet via \texttt{DELEGATECALL} \cite{Explained:2024}. In ByBit (2025), attackers pushed a forged implementation contract into the exchange's cold-wallet proxy, overwriting storage and seizing ownership by abusing Safe's upgrade path \cite{bybit_certik} (see \autoref{sec:bybit_case}). The classic Parity library bug (2017) involved an uninitialised contract that allowed the adversary to gain ownership and drain multi-sig wallets \cite{palladino2017parity}. These cases map to two recurrent sub-patterns: \begin{enumerate*}[label=(\arabic*)] \item upgrade-path hijack, where the authorised proxy-upgrade or module-installation channel is abused to introduce attacker-controlled logic (ByBit, WazirX); and \item constructor hijack, where the \texttt{init} function is left callable after deployment (Parity). \end{enumerate*}

\subsection{Authentication}
\label{sec:auth-attacks}

\subsubsection{Credential Cracking}
\label{sec:cred-crack}

This category of attacks systematically attempts different credential values to bypass the authentication mechanism. Brute-force attacks involve an adversary systematically trying all possible character combinations to bypass the authentication function and decrypt \textcolor{teal}{\textit{enc\_sk}}. If successful, the adversary can create malicious transactions using \hyperref[algo:transaction-signing]{Algorithm 2} \cite{Kiktenko2019DetectingWallets}. Dictionary attacks, on the other hand, leverage commonly used words to predict \textcolor{teal}{\textit{rdm\_seed}} phrases for access. Unlike brute-force attacks that exhaust all possible combinations, dictionary attacks are computationally less demanding, and their success rate increases with the use of leaked password datasets \cite{Uddin2021Horus:Wallets, praitheeshan2019security}.

\subsubsection{Identity Spoofing}
\label{sec:iden-spoof}

For enhanced \acs{kek} security, wallets leverage supplementary user authentication methods, such as user biometrics and \acf{2fa} implementations.

The identity spoofing attack method bypasses these verification mechanisms (see \hyperref[algo:cryptocurrency-wallet]{Algorithm 1}) by impersonating the user to decrypt \textcolor{teal}{\textit{enc\_sk}} and authorise malicious transactions. In fake biometric attacks, an adversary employs synthetic or reconstructed biometric data to achieve this goal \cite{galbally2013image}. To circumvent SMS-based \acs{2fa}, an adversary can also use SIM swap attacks, which execute the transfer of the user's phone number to an adversary-controlled mobile device \cite{Kim2022ACountermeasures}. Mobile wallets, smart contract wallets and other infrastructures that integrate SMS-based \acs{2fa} or biometric verification can be vulnerable to these attacks (see \autoref{tab:attack_vectors}).

\subsection{Storage and Memory}
\label{sec:physical-attacks}

\subsubsection{Physical Tampering}
\label{sec:tam-per}

These primarily involve physically altering a wallet’s hardware to bypass security protections. In evil maid attacks, the attacker physically modifies the unencrypted storage of a device to capture credentials or manipulate the system \cite{altuwaijri2020android}. In contrast, microscopy attacks use advanced techniques, such as electron microscopy, to examine the microelectronic components of a wallet. These attacks can extract critical data or identify vulnerabilities, often without altering the hardware itself \cite{courbon2016reverse}.

\subsubsection{Fault Injection}
\label{sec:fau-inj}

These attacks manipulate the wallet's components by forcing an erroneous system state to bypass the security mechanisms \cite{Akter2023AChallenges}. For instance, fault injection attacks on hardware wallets often exploit vulnerabilities in volatile memory (such as \acs{sram}) by manipulating environmental factors. Data remanence vulnerabilities in the Trezor wallet have been exploited to demonstrate these attacks \cite{trezor_memory, trezor_medium}. Fault injection attacks on smart contracts have also been shown in the literature \cite{hajdu2020using}.

\subsubsection{Other Non-Invasive Techniques}
\label{sec:non-inv-man}

Other non-invasive storage and memory attacks exist which are not based on fault injection methods. In cold boot attacks, the attacker executes a cold restart on the wallet device to exploit the data remanence properties of volatile memory, such as \acf{dram} and \acf{sram}, to retrieve sensitive data \cite{Shaikh2022SurveyExchanges}. Similarly, \acs{puf} attacks exploit the unique characteristics of hardware defence implementations known as \acf{puf}. These implementations have challenge-response functionality that exhibits physical unclonability \cite{Garcia-Bosque2020IntroductionApplications, wang2024efficient}.


\subsection{Cryptanalysis}
\label{sec:cryptanalysis-analysis}


\subsubsection{Side-Channel Analysis}
\label{sec:side-channel}

Non-invasive key extraction attacks on cryptographic functions, including timing and power \acf{sca}, are executed by exploiting side channels. These attacks exploit leakages in behaviours exhibited by cryptographic functions (see \autoref{sec:wallet_mechanism}) through side channels to measure and extract values such as time and power  \cite{Shaikh2022SurveyExchanges, Park2023}. Timing-based \acs{sca} measures the cryptographic function execution time. Successful implementation of a timing-based side-channel attack has been demonstrated on a Trezor One hardware wallet \cite{kocher1996timing}. Power-based \acs{sca} analyses the cryptographic function's power trace, including the hash function. \acs{sca} on the hash function has been utilised to extract the \textcolor{teal}{\textit{rdm\_seed}} \cite{Park2024CloningFunction}.

\subsubsection{Direct Exploitation}
\label{sec:impl-exp}

These attacks directly target implementation errors within the cryptographic surface area. Weak signature (\teal{$\sigma$}) attacks, for example, target weaknesses in the signing algorithm due to improper implementation, weak or outdated cryptographic algorithms or errors in encryption logic \cite{Rokhjavan2023SecuringWallets}. In addition, an adversary can exploit vulnerabilities in \hyperref[algo:transaction-signing]{Algorithm 2} by reusing a nonce during transaction authorisation \cite{brengel2018identifying}. Such reuse can compromise the security of wallets by resulting in \textcolor{teal}{\textit{sk}} leakage \cite{Ko2020PrivateSignatures}.

\section{Security Measures}
\label{sec:defense-strategies}

This section builds upon the framework outlined in \autoref{sec:attack-framework} by presenting mitigation approaches against wallet attacks. We aim to examine defence mechanisms for each identified attack vector affecting wallets.

\subsection{Network}
\label{sec:net-def}

Suspicious network activity can be detected through machine learning techniques, including anomaly detection models \cite{kapoor2021ransomware} and classification algorithms \cite{balakrishnan2023analysis}. Additionally, dynamic network parameter adjustments \cite{Girdler2021ImplementingAddresses} and other intrusion detection mechanisms \cite{guri2018beatcoin, zimba2019cryptojacking} further contribute to identifying such anomalies.

\begin{landscape}
\begin{table}[!htbp]
\centering
\caption{
Three-level attack classification showing gap analysis, threat occurrences, adversary's target and mapping to possible security measures (\autoref{sec:defense-strategies}). The \enquote{Gaps} summary shows that academic literature covers 24 of the 28 enumerated attack vectors (86\%), whereas publicly reported incidents cover 9 vectors (32\%). Notable incident percentages are calculated from a total of 85 reported industry incidents (see \autoref{tab:attack-incidents}). Symbols: ( \smallfullcirc : include, \smallhalfcirc : part-inclusion (influenced by other factors), \smallemptycirc : not include) }
\label{tab:attack_vectors}
\tiny
\renewcommand{\arraystretch}{1}
\setlength{\tabcolsep}{1.5pt} 
\resizebox{\linewidth}{!}
{
\begin{tabular}{llllccccccccccccccccccccccccccccccccccccccccccccccccl}
\toprule
\multicolumn{1}{c}{\textbf{Category}} &
  \multicolumn{1}{c}{\multirow{1}{*}{\textbf{\hyperref[sec:attack-framework]{ \textbf{Method}}}}} &
  \multicolumn{1}{c}{\multirow{1}{*}{\textbf{\hyperref[sec:attack-framework]{\textbf{Vector}}}}} &
  \multicolumn{21}{c}{\textbf{\hyperref[sec:threat_framework]{Threat}}} &
  \multicolumn{16}{c}{\textbf{\hyperref[sec:wallet_mechanism]{Target}}} & 
  \multicolumn{3}{c}{\textbf{\hyperref[sec:adversary_goal]{Goal}}} &
  \multicolumn{5}{c}{\textbf{\hyperref[sec:infrastructure]{Infrastructure}}} &
  \multicolumn{2}{c}{\textbf{\hyperref[sec:attacks_discussion]{Gaps}}} &
  \multicolumn{1}{c}{\textbf{\hyperref[sec:defense-strategies]{Possible Defence}}} \\
\bottomrule
\end{tabular}}
\end{table}
\clearpage
\end{landscape}

To mitigate these attacks, wallets can adopt network security protocols that validate and authenticate IP addresses \cite{rengarajan2016secure} and incorporate additional security layers within the wallet's network to prevent potential \teal{$txn$} modification attempts by adversaries \cite{Cai2014ResearchNetwork}. To limit or prevent \acf{ddos} attacks, wallets must distinguish malicious and authentic network traffic using classifiers such as the decision tree algorithm \cite{khan2019adaptive} and reinforcement learning approaches to analyse patterns in network data \cite{liu2018deep}. Another mitigation approach involves analysing the network for unusual patterns, such as repeated request attempts from the same \acs{ip} address \cite{sathwara2017distributed}.

\subsection{Application}
\label{sec:app-def}

To mitigate the risk of message alteration by clipboard hijackers, wallets can employ features such as NFC and two-dimensional codes to prevent recipient address modification during transaction creation \cite{li2020android}. From a user perspective, human-readable addresses such as \acs{ens} \cite{ENS2024EthereumService} aid in detecting address tampering, though they have certain security vulnerabilities \cite{Xia2022ChallengesENS}. Wallets can also prevent system behaviour modifications by addressing specific attack vectors. Attack vectors that attempt these modifications by targeting vulnerabilities in the \acs{os} can be mitigated by employing code obfuscation \cite{indusface} and runtime protection mechanisms \cite{qi2012spad}. Furthermore, by enforcing \acf{cfi} measures, wallets can ensure that control flow cannot be hijacked to deviate from intended control flow paths for malicious transactions \cite{Creech2017NewMitigation}. 

\subsection{Authentication}
\label{sec:auth-def}

Wallets can incorporate features either as direct protection against specific attack methods or as general authentication bypass protection. By directly integrating improved functionalities to obstruct access to predictive text data, wallets can prevent dictionary attacks \cite{Uddin2021Horus:Wallets}. Additionally, to prevent brute-force attacks, only complex passwords should be allowed in the initialisation stage  \cite{praitheeshan2019security}. Biometric falsifying attacks can be prevented by incorporating liveness detection features in wallets \cite{galbally2013image}.

To prevent single points of failure, wallets can enhance authentication levels (\autoref{sec:design-authen}) through \acs{mfa}, \acf{mpc} \cite{Lindell2020SecureComputation} and multi-signatory features such as \acs{bip}-11's M-of-N standard  \cite{bip11} (\autoref{sec:design-distr}). To mitigate social engineering attacks, wallets can incorporate phishing-resistant \acs{mfa} techniques such as FIDO2 \cite{Wang2021OnAttacks}. This feature enables communication with the original wallet website to verify authenticity before allowing access to the wallet \cite{fido2}.

\subsection{Storage and Memory}
\label{sec:sto-def}

An effective defence method against these attacks involves incorporating \acf{puf} to generate cryptographic keys on demand, without storing \teal{$sk$} on the wallet's chip. This method also prevents microscopy attacks, some other physical tampering attacks, and side-channel attacks (see \autoref{sec:crypt-def}) \cite{Urien2021InnovativeWallets, Park2024CloningFunction}. Physical tampering through the evil maid attack can be limited by implementing trusted boot mechanisms \cite{Tereshkin2010EvilEncryption}. Possible mitigations against non-invasive manipulation, such as the cold boot attack, involve adopting features which algorithmically clear the wallet's memory following intrusion \cite{seol2019amnesiac}. For example, Ledger has introduced a secure layer which detects chip intrusion and erases \teal{$sk$} following extraction attempts \cite{ledgerwallet}.

\begin{table*}[!h]
\centering
\renewcommand{\arraystretch}{1.1}
\setlength{\tabcolsep}{1.5pt} 
\footnotesize 
\resizebox{1.0\textwidth}{!}{
\begin{tabular}{llcccccccccccccccccccccccccccccc}
\toprule
\vspace{1pt} 
& \multicolumn{31}{c}{\textbf{Possible Defence Methods}}
\vspace{1pt} 
\\
\multicolumn{2}{c}{\textbf{ Classification}} 
& \rotatebox[origin=c]{90}{\cite{Cai2014ResearchNetwork}} 
& \rotatebox[origin=c]{90}{\cite{Ahmed2017MitigatingNetworking}} 
& \rotatebox[origin=c]{90}{\cite{Bhirud2011LightPrevention}} 
& \rotatebox[origin=c]{90}{\cite{liu2018deep}} 
& \rotatebox[origin=c]{90}{\cite{sathwara2017distributed}} 
& \rotatebox[origin=c]{90}{\cite{li2020android}} 
& \rotatebox[origin=c]{90}{\cite{ferdous2023review}} 
& \rotatebox[origin=c]{90}{\cite{indusface}} 
& \rotatebox[origin=c]{90}{\cite{Tirronen2018StoppingData}} 
& \rotatebox[origin=c]{90}{\cite{Aratani2015AuthenticationChannel}} 
& \rotatebox[origin=c]{90}{\cite{aldawood2020advanced}} 
& \rotatebox[origin=c]{90}{\cite{galbally2013image}} 
& \rotatebox[origin=c]{90}{\cite{altuwaijri2020android}} 
& \rotatebox[origin=c]{90}{\cite{breier2022practical}} 
& \rotatebox[origin=c]{90}{\cite{Urien2021InnovativeWallets}} 
& \rotatebox[origin=c]{90}{\cite{Gupta2019ImpactSecurity}} 
& \rotatebox[origin=c]{90}{\cite{brengel2018identifying}} 
& \rotatebox[origin=c]{90}{\cite{Park2024CloningFunction}} 
& \rotatebox[origin=c]{90}{\cite{Akter2023AChallenges}} 
& \rotatebox[origin=c]{90}{\cite{Lindell2020SecureComputation}} 
& \rotatebox[origin=c]{90}{\cite{bip11}} 
& \rotatebox[origin=c]{90}{\cite{Park2023}} 
& \rotatebox[origin=c]{90}{\cite{Feng2023Man-in-the-middleRedirects}} 
& \rotatebox[origin=c]{90}{\cite{Kim2022ACountermeasures}} 
& \rotatebox[origin=c]{90}{\cite{Shuvo2023AAttacks}} 
& \rotatebox[origin=c]{90}{\cite{zimba2019cryptojacking}} 
& \rotatebox[origin=c]{90}{\cite{qi2012spad}} 
& \rotatebox[origin=c]{90}{\cite{ManageAddresses}} 
& \rotatebox[origin=c]{90}{\cite{hu2020overview}} 
& \# (\%)
\vspace{1.5pt} 
\\
\midrule
\multirow{3}{*}{Precautionary} &
\rotatebox[origin=c]{0}{Prevention} & {\smallemptycirc} & {\smallemptycirc} & {\smallemptycirc} & {\smallemptycirc} & {\smallemptycirc} & {\smallfullcirc} & {\smallemptycirc} & {\smallemptycirc} & {\smallfullcirc} & {\smallemptycirc} & {\smallemptycirc} & {\smallemptycirc} & {\smallemptycirc} & {\smallemptycirc} & {\smallemptycirc} & {\smallemptycirc} & {\smallfullcirc} & {\smallemptycirc} & {\smallemptycirc} & {\smallemptycirc} & {\smallemptycirc} & {\smallemptycirc} & {\smallemptycirc} & {\smallemptycirc} & {\smallemptycirc} & {\smallemptycirc} & {\smallemptycirc} & {\smallemptycirc} & {\smallemptycirc} & \cellcolor{g2}{$3$($10\%$)} \\
& \rotatebox[origin=c]{0}{Protection} & {\smallfullcirc} & {\smallfullcirc} & {\smallfullcirc} & {\smallemptycirc} & {\smallemptycirc} & {\smallfullcirc} & {\smallfullcirc} & {\smallfullcirc} & {\smallemptycirc} & {\smallfullcirc} & {\smallfullcirc} & {\smallfullcirc} & {\smallemptycirc} & {\smallemptycirc} & {\smallfullcirc} & {\smallfullcirc} & {\smallemptycirc} & {\smallfullcirc} & {\smallfullcirc} & {\smallemptycirc} & {\smallemptycirc} & {\smallfullcirc} & {\smallfullcirc} & {\smallemptycirc} & {\smallemptycirc} & {\smallemptycirc} & {\smallfullcirc} & {\smallemptycirc} & {\smallfullcirc} &  \cellcolor{g6}{$17$($58\%$)} \\
& \rotatebox[origin=c]{0}{Limitation} & {\smallemptycirc} & {\smallemptycirc} & {\smallemptycirc} & {\smallfullcirc} & {\smallemptycirc} & {\smallemptycirc} & {\smallemptycirc} & {\smallemptycirc} & {\smallemptycirc} & {\smallemptycirc} & {\smallemptycirc} & {\smallemptycirc} & {\smallfullcirc} & {\smallemptycirc} & {\smallemptycirc} & {\smallemptycirc} & {\smallemptycirc} & {\smallemptycirc} & {\smallemptycirc} & {\smallfullcirc} & {\smallfullcirc} & {\smallemptycirc} & {\smallemptycirc} & {\smallfullcirc} & {\smallemptycirc} & {\smallemptycirc} & {\smallemptycirc} & {\smallfullcirc} & {\smallemptycirc} & \cellcolor{g3}{$6$($21\%$)} \\
\midrule
\multirow{3}{*}{Remedial} & \rotatebox[origin=c]{0}{Detection} & {\smallemptycirc} & {\smallemptycirc} & {\smallemptycirc} & {\smallfullcirc} & {\smallemptycirc} & {\smallemptycirc} & {\smallemptycirc} & {\smallemptycirc} & {\smallemptycirc} & {\smallemptycirc} & {\smallemptycirc} & {\smallemptycirc} & {\smallemptycirc} & {\smallfullcirc} & {\smallemptycirc} & {\smallemptycirc} & {\smallemptycirc} & {\smallemptycirc} & {\smallemptycirc} & {\smallemptycirc} & {\smallemptycirc} & {\smallemptycirc} & {\smallfullcirc} & {\smallemptycirc} & {\smallfullcirc} & {\smallfullcirc} & {\smallemptycirc} & {\smallemptycirc} & {\smallemptycirc} & \cellcolor{g3}{$5$($17\%$)} \\
& \rotatebox[origin=c]{0}{Response} & {\smallemptycirc} & {\smallemptycirc} & {\smallemptycirc} & {\smallfullcirc} & {\smallemptycirc} & {\smallemptycirc} & {\smallemptycirc} & {\smallemptycirc} & {\smallemptycirc} & {\smallemptycirc} & {\smallemptycirc} & {\smallemptycirc} & {\smallemptycirc} & {\smallemptycirc} & {\smallemptycirc} & {\smallemptycirc} & {\smallemptycirc} & {\smallemptycirc} & {\smallemptycirc} & {\smallemptycirc} & {\smallemptycirc} & {\smallemptycirc} & {\smallemptycirc} & {\smallemptycirc} & {\smallemptycirc} & {\smallemptycirc} & {\smallemptycirc} & {\smallemptycirc} & {\smallemptycirc} & \cellcolor{g1}{$1$($3\%$)} \\
& \rotatebox[origin=c]{0}{Recovery} & {\smallemptycirc} & {\smallemptycirc} & {\smallemptycirc} & {\smallemptycirc} & {\smallfullcirc} & {\smallemptycirc} & {\smallemptycirc} & {\smallemptycirc} & {\smallemptycirc} & {\smallemptycirc} & {\smallemptycirc} & {\smallemptycirc} & {\smallemptycirc} & {\smallemptycirc} & {\smallemptycirc} & {\smallemptycirc}  & {\smallemptycirc} & {\smallemptycirc} & {\smallemptycirc} & {\smallemptycirc} & {\smallemptycirc} & {\smallemptycirc} & {\smallemptycirc} & {\smallemptycirc} & {\smallemptycirc} & {\smallemptycirc} & {\smallemptycirc} & {\smallemptycirc} & {\smallemptycirc} & \cellcolor{g1}{$1$($3\%$)}
\vspace{1pt}
\\
\midrule 
\multicolumn{3}{c}{Summary}  &
\multicolumn{8}{c}{Precautionary:  \cellcolor{g6}{$26$($89\%$)}}  &
\multicolumn{8}{c}{Remedial: 
 \cellcolor{g3}{$7$($24\%$)}}  &
\multicolumn{11}{r}{Total Unique Methods    }  &
\multicolumn{1}{c}{}  &
\cellcolor{g0}{$29$($100\%$)} 
\vspace{1pt} 
   \\
\bottomrule
\end{tabular}
}
\vspace{1ex} 
\caption{Defence methods categorised by type showing classification frequency (\#) and percentage (\%). Precautionary methods proactively prevent attacks; remedial methods provide attack detection, response, or data recovery.}
\label{tab:defence_methods}
\end{table*}



\subsection{Cryptanalysis}
\label{sec:crypt-def}

Exploiting cryptographic vulnerabilities can lead to \teal{$sk$} extraction. Attacks that aim to exploit weak cryptographic signatures (\teal{$\sigma$}) can be counteracted by employing stronger hashing algorithms \cite{Rokhjavan2023SecuringWallets}, while deterministic \teal{$nonce$} selection prevents nonce reuse attacks \cite{brengel2018identifying}. Non-invasive attacks on cryptographic functions, including timing and power \acs{sca}, are executed by exploiting side channels. Effective prevention methods include data leakage protection and disguising data access patterns as noise injection \cite{Akter2023AChallenges, Lou2021ACryptography, Ali2023CharacterizationHardware, Park2024CloningFunction}. These affect the adversary's ability to interpret leaked information effectively \cite{Mosquera2023GuardAttacks}.

\section{Case Studies}
\label{sec:case_study}

In this section, we present detailed case studies of notable wallet security breaches. We apply our wallet design taxonomy (\autoref{sec:wallet-taxonomy}), threat model (\autoref{sec:threat_framework}), and attack taxonomy (\autoref{sec:attack-framework}). Each case study systematically analyses the wallet's architecture, identifies exploited vulnerabilities, and explores the sequence of attack events. We conclude each study with recommended and implemented security measures.

\subsection{Case Study: ByBit Custodial Wallet Hack}
\label{sec:bybit_case}

In February 2025, ByBit experienced a significant security breach that resulted in a loss of approximately \$1.5 billion in Ethereum, marking the largest cryptocurrency theft to date \cite{bybit}. This sophisticated attack aligns with the attack vectors outlined by our taxonomy. We provide a detailed analysis below using our frameworks for design classification, threat assessment, attack sequence analysis, and mitigation strategies.

\subsubsection{Design}
\label{sec:bybit_mech}

Using our design taxonomy in \autoref{sec:wallet-taxonomy}, we analyse the ByBit wallet design below:

\begin{itemize}
    \item \textbf{Custody:} ByBit maintained full custody of user funds, with users relinquishing \textcolor{teal}{\textit{sk}} control to the exchange. This particular case pertains to the \textcolor{teal}{\textit{sk}}, which controlled the Ethereum assets of the exchange. 
    \item \textbf{Infrastructure:} 
    ByBit employed a multi-faceted infrastructure design, integrating hardware wallets with a smart contract-enabled proxy architecture. The primary proxy contract delegated logic execution to a separate implementation contract via \texttt{delegateCall}. It stored the implementation contract's address in storage slot 0 to facilitate future upgrades \cite{bybit_secux}. However, the design did not enforce strict access controls on this critical operation. This became a key factor exploited in the attack, as described in the threat analysis (see \autoref{sec:bybit_dep}).

    \item \textbf{Distribution:} \textcolor{teal}{\textit{sk}} management was distributed securely with authorisation rights shared among multiple private key (\textcolor{teal}{\textit{sk}}) holders in the multi-sig scheme across different hardware devices. The multi-signature scheme prevented unilateral transactions, mandating consensus among multiple trusted individuals. 
    \item \textbf{Authorisation:} Transactions were generated via Safe's web interface. Signers reviewed transaction details on the web user interface and hardware wallet screens. Only after confirmation on their Ledger hardware wallet devices were transactions broadcast to the blockchain.
    \item \textbf{Validation:} After obtaining the necessary approvals, transactions underwent validation to ensure compliance with ByBit's internal security policies. This included verifying adherence to address whitelisting protocols and transfer limits. The multi-sig smart contract enforced these policies by executing transactions only when the requisite number of valid signatures was present.
\end{itemize}

\subsubsection{Threats and Dependencies}
\label{sec:bybit_dep}

ByBit’s security architecture relied significantly on several interconnected elements, including the Safe user interface, which proved vulnerable to the adversaries' attempts. We outline the threats, which were exploited by the adversary inline with our threat model below:

\begin{itemize}
    \item \textbf{Insecure Interaction:} Insecure interactions resulted in the system's exposure to threats. The adversary likely exploited these interactions to achieve infiltration of the Safe developer's machine \cite{bybit_certik}. 
    \item \textbf{Application Provider Compromise:} ByBit's operational security was heavily dependent on the integrity and security posture of third-party service providers, in this case, Safe’s web interface.
    \item \textbf{Data Misrepresentation:} The adversary compromised the accuracy and reliability of transaction data presented to authorised signers through Safe's user interface. This highlighted a critical vulnerability in wallet user interfaces.
    \item \textbf{Application Logic Flaw:} The infrastructure design permitted unrestricted use of the \texttt{delegateCall} instruction, allowing malicious actors to overwrite critical storage slots. Specifically, the attackers exploited the ability to overwrite the logic pointer stored in storage slot 0, leading to unauthorised control of the proxy's logic \cite{bybit_certik}. This violated the principle of least privilege and directly facilitated the privilege escalation step of the attack.
    \item \textbf{Blind Signing:} ByBit's reliance on hardware wallet confirmation processes did not sufficiently address the blind signing risk. Signers assumed the hardware wallet displays were a trustworthy verification source and approved transactions without explicit visibility into critical transaction metadata. This included \texttt{delegateCall} operations and underlying implementation changes. 

\end{itemize}

\subsubsection{Adversary Goal and Capabilities}
\label{sec:bybit_cap}

\textcolor{teal}{\textit{A}} aimed to gain unauthorised rights by masking adversary-created transactions as benign. The capabilities of \textcolor{teal}{\textit{A}} significantly evolved during the attack as extended knowledge was gained, starting from restricted external knowledge and progressing to insider-level knowledge and access:

\begin{itemize}
    \item \textbf{Initial Phase:} \textcolor{teal}{\textit{A}} remotely exploited publicly accessible information to exploit Safe developer interactions and gain restricted internal access.
    \item \textbf{Intermediate Phase:} Having achieved insider-level knowledge and privileges following a successful repository compromise, \textcolor{teal}{\textit{A}} could inject malicious software into operational components of the wallet software.
    \item \textbf{Final Phase:} \textcolor{teal}{\textit{A}} could exploit application logic to deceive \textcolor{teal}{\textit{sk}} holders, achieving credential compromise. Subsequently, \textcolor{teal}{\textit{A}} gained full wallet control and authorisation rights.
\end{itemize}

\subsubsection{Attack Sequence}
\label{sec:bybit_att}

The ByBit incident represents a sophisticated combination of several coordinated attack vectors identified in our Application threats taxonomy:

\begin{itemize}
    \item \textbf{Social Engineering:} A phishing attack method enabled the execution of subsequent attack vectors. Social engineering and malware were combined to compromise ByBit, as seen in past incidents (e.g., BitKeep \cite{CertiKIncidents}, Upbit \cite{UpbitMedium}, and wallet drainers \cite{RektREKT}). This gave the adversary direct access to Safe's front-end code repository, highlighting the importance of secure developer environments.
    \item \textbf{Malware Execution:} The compromised machine enabled the injection of malicious JavaScript into Safe's front-end code, targeting the transaction approval interface. The malware modified the transaction data displayed to \textcolor{teal}{\textit{sk}} holders. While legitimate transaction details were displayed in the Safe wallet user interface, the data sent to the hardware wallet was altered.
    \item \textbf{Privilege Escalation:} 
    The approved transaction altered the smart contract's logic. The attackers exploited storage slot hijacking by crafting a transaction that used \texttt{delegateCall} to execute a spoofing contract. This contract’s \texttt{transfer()} function wrote the attacker’s malicious implementation address to storage slot 0 via the \acf{evm} \texttt{SSTORE} opcode, overwriting the proxy’s logic pointer. With the proxy now delegating to the attacker’s contract, all subsequent transactions executed attacker-controlled code in the proxy’s context, granting full authorisation rights.

\end{itemize}

\subsubsection{Security Measures}
\label{sec:bybit_def}

Before the breach, ByBit used a layered security model: most funds were in a Safe contract, private keys on six Ledger devices, requiring 4-of-6 multi-sig. These measures were bypassed. After the incident, industry experts highlighted the following additional controls:

\begin{itemize}
    \item \textbf{Independent Transaction Hash Verification:} The use of tools such as \texttt{safe-tx-hashes} to independently verify transaction hashes against on-chain data mitigates the risk of UI-level deception \cite{bybit_cyfrin}. By enabling signers to cross-reference actual transaction payloads outside of potentially compromised interfaces, this approach detects malicious operations such as unauthorised \texttt{delegateCall} or logic pointer overwrites before execution.
    \item \textbf{Transaction Policy Enforcement via On-Chain Gatekeeping:} Preventative solutions such as Halborn’s Seraph simulate signed transactions before execution and block operations that violate predefined organisational policies \cite{bybit}. In the context of the ByBit attack, this approach could have flagged and halted the unauthorised upgrade triggered by the malicious \texttt{delegateCall}, enforcing a secondary layer of validation beyond signer intent.
    \item \textbf{Hardware Wallet Clear-Signing:} Require devices that support the on-device display of the complete destination, value, function selector, and raw calldata (clear-signing) before approval. This enables signers can independently verify every field and avoid hash-only blind signing, a weakness exploited in the ByBit breach \cite{bybit_secux}.    
    \item \textbf{Wallet Auditing:} Conducting regular audits focusing on storage layout consistency and \texttt{delegateCall} whitelisting and other wallet-related code is pertinent \cite{bybit_slowmist}
\end{itemize}

\subsection{Case Study: Slope Non-Custodial Wallet Hack} \label{sec:slope_case}

In August 2022, Slope Wallet experienced a severe security incident, resulting in the compromise of over 9,200 user wallets on the Solana blockchain and a loss of approximately \$4.1 million in SOL and USDC \cite{CoinTelegraph2022SlopeAttack}. We provide a detailed analysis below using our frameworks for design classification, threat assessment, attack sequence analysis, and implemented security measures.

\subsubsection{Design} \label{sec:slope_design}

Applying our design taxonomy, we analyse the Slope wallet design below:

\begin{itemize} 
\item \textbf{Custody:} Slope utilised a non-custodial model where users retained complete control over the private key (\textcolor{teal}{\textit{sk}}). This case pertains to the management and leakage of the user's private key.
\item \textbf{Infrastructure:} Slope used a mobile software wallet that relied on a self-hosted Sentry monitoring stack \cite{cyberintel_slope, fyeo_slope}. This setup collected application data for debugging but inadvertently logged sensitive information due to a faulty logging function.
\item \textbf{Distribution:} Slope used a single-distribution model, with all cryptographic operations and storage conducted solely on the user’s mobile device. No advanced key distribution methods, such as MPC or multi-signature schemes, were integrated. 
\item \textbf{Authorisation and Validation:} The standard Solana \textit{Ed25519} signature flow was executed locally on the user device. Transaction broadcasting was performed via Slope's own \acs{rpc} endpoints.
 \end{itemize}

\subsubsection{Threats and Dependencies} \label{sec:slope_dep}

Slope’s security architecture relied heavily on interconnected dependencies, particularly its integrated application-monitoring stack, as detailed below:
\begin{itemize} \item \textbf{Application‑Monitoring Dependency:} Slope utilised an on-premise implementation of the Sentry SDK, designed to assist developers in debugging. A single improperly added \texttt{toString()} method circumvented built-in security filters, resulting in sensitive wallet private keys being unintentionally logged in plaintext \cite{cyberintel_slope}.
\item \textbf{Data Leakage:} Multiple defensive measures were used (collection filtering, \acf{tls} certificate pinning, database encryption at rest). However, collection filtering and database encryption were disabled, causing plaintext private keys to be stored in the database.
\item \textbf{Third‑Party Supply‑Chain Threat:} Slope employed a self-hosted version of the third-party monitoring solution (Sentry), inheriting risks associated with configuration drift, patch management latency, and internal operational errors. This on-premise deployment introduced vulnerabilities typically mitigated by a SaaS-managed setup.
\item \textbf{Insecure User Interaction:} Users continued to interact with wallets whose keys had potentially been exfiltrated. No built-in key-rotation prompt existed. \end{itemize}

\subsubsection{Adversary Goal and Capabilities} \label{sec:slope_cap}

The adversary, \textcolor{teal}{\textit{A}}, aimed primarily for credential compromise, specifically targeting the user's private key (\textcolor{teal}{\textit{sk}}). The capabilities leveraged by \textcolor{teal}{\textit{A}} included:

\begin{itemize} 
\item \textbf{Initial Phase:} \textcolor{teal}{\textit{A}} used knowledge of Slope’s logging vulnerability (via reverse engineering or insider information) to target the timeframe and method to extract logged private keys. 
\item \textbf{Intermediate Phase:} \textcolor{teal}{\textit{A}} employed remote network access, either directly to the internal database or by intercepting \acs{tls} traffic prior to 18 July 2022. This remote capability allowed the extraction of plaintext private keys despite the security measures initially in place.
\item \textbf{Final Phase:} \textcolor{teal}{\textit{A}} employed legitimate wallet signing authority using stolen keys and subsequently drained user funds directly via standard blockchain transactions without triggering conventional anomaly detection.
\end{itemize}

\subsubsection{Attack Sequence} \label{sec:slope_att}

In this incident, the adversary employed the logic exploitation vector to compromise credentials, as summarised below:

\begin{itemize} 
\item \textbf{Logic Bug Introduction:} Slope utilised a helper function (\texttt{toString()}) to streamline debugging, unintentionally bypassing established security filters. This bug directly caused private keys to enter plaintext logging pipelines.
\item \textbf{Data Pipeline Restart:} Slope utilised Kafka for data processing. After restarting, Kafka inadvertently flushed cached logs containing private keys in plaintext format directly into a PostgreSQL database.
\item \textbf{Log Exfiltration:} \textcolor{teal}{\textit{A}} accessed the misconfigured Sentry instance and retrieved the plaintext seed phrases, fully compromising user private keys.
\item \textbf{Wallet draining:} \textcolor{teal}{\textit{A}} utilised legitimate signing authority gained from compromised private keys and drained assets from 9,229 wallet addresses within seven hours.

\end{itemize}

\subsubsection{Security Measures
} 
\label{sec:slope_def}

Following the Slope Wallet breach, the development team initiated several immediate reactive security measures. The team promptly disabled the self-hosted Sentry server within 15 minutes of identifying the vulnerability and advised users to transfer their assets to new wallets \cite{dailycoin_slope}. Additionally, audits conducted by \href{https://osec.io/}{OtterSec} and \href{https://www.slowmist.com/}{Slowmist} confirmed that sensitive data, including private keys, had been inadvertently logged \cite{ottersec_slope, zellic_slope}. In response, Slope removed all sensitive logging functionalities and implemented a \texttt{beforeSend} whitelist to filter out confidential information \cite{slope_statement}. 

To prevent such incidents in the future, it is crucial to ensure that application monitoring tools, such as Sentry, are meticulously configured to exclude sensitive data from logs. This involves implementing stringent data scrubbing protocols and avoiding the logging of private keys or seed phrases. Proper calibration of these safeguards is essential for preserving the confidentiality and integrity of user credential data.

\section{Insights}
\label{sec:insights}

We discuss insights on design, threats, attack methods, and security measures from academic papers, industry incidents, and case studies below:

\subsection{Influence of Design on Threats}
\label{sec:threats_dis_influence}

Despite a wide range of security setups, we observe that the majority of the design combinations of existing wallets surveyed have been threatened by multiple vulnerabilities, as shown in \autoref{tab:wlt._taxonomy}. This is due to similar implementations i.e., the use of replicated libraries and commonly integrated implementation proposals (e.g., \acs{eip}-4337). We also observe that some wallets have had numerous vulnerabilities discovered in industry and academia. Most notably, Ledger and Trezor have several data remanence, data manipulation and insecure cryptographic vulnerabilities. Furthermore, in mapping vulnerabilities to attacks, we observe that some vulnerabilities can lead to numerous attack vectors as shown in \autoref{fig:wallet-mapping}. These include inadequate authentication, insecure permissions, insecure user interactions, and particularly data leakage. The Slope Wallet incident exemplifies this, where an improperly configured debug logging mechanism led directly to private key leakage.

\subsection{High Occurrence of Signature Verification Logic Flaws}
\label{sec:sig_verif_flaw}

We observe that signature verification logic flaws account for the most vulnerability occurrences in various wallets surveyed, constituting 19\%. Another interesting observation is the occurrence of this vulnerability in three diverse wallet security enhancement architectures, namely hardware, smart contract and \acs{mpc} wallets \cite{cve_14199, fireblocks_23, AccountMedium, UncoveringVulnerability}.

\subsection{Gap Analysis on Wallet Threats}

Conducting a gap analysis across industry and academic reports is difficult because many incidents do not disclose precise attack methods. We generally observe a high correlation between identified threats in industry and academia, except for insider and external threats. Specifically, in the following threats: malicious insider, compromised insider and compromised service provider threats. Although several custodial designs have been proposed by academia along with threat models, an investigation into the potential external threats and attacks in custodial setups would be highly beneficial for the industry. Notably, most industry attacks target exchanges and other custodial setups, as large funds are concentrated within a few wallet addresses. Additionally, research into these areas will also be pertinent due to the fact that wallet designs are gradually evolving into shared-custodial or other setups which require authentication from a centralised party (e.g., passkey, \acs{2fa}).

To address the gaps identified in \autoref{tab:threat_capability}, we propose the following measures:  
\begin{itemize}
    \item \textbf{Responsible Disclosure Policies:} Create a standardised incident template for responsible disclosure of wallet-related incidents. This could employ a uniform reporting format for exchanges and custodians to use when disclosing incidents, enabling both industry and academic audiences to analyse them consistently. A notable example in industry is Immunefi’s vulnerability disclosure platform \cite{immunefi2024}.
    
    \item \textbf{Public-Private Collaborations:} Formalise partnerships between exchanges, blockchain security firms, and academic institutions to analyse incident data. Successful models exist, including as IC3 and Chainlink partnership \cite{ic3} and the Stanford Centre for Blockchain Research’s industry partnerships \cite{stanford_cbr}.

       \item \textbf{Open-source Incident Registry:} Develop an open repository where vetted blockchain incident post-mortems can be deposited by operators and accessed by researchers, policymakers, and other exchanges. An existing example is the SlowMist Hacked incident archive \cite{slowmist_hacked}.

\end{itemize}

\subsection{Difference in Academia and Notable Industry Incidents}

Identifying attack vectors within the industry remains challenging, as sources often lack specificity. Notable attack vectors are significantly less clear (46\% unknown) and show a lower spread compared to attacks described in the literature (see \autoref{tab:attack_vectors}). This might be attributed to a lack of detailed post-mortem analysis in several incidents and an adversary's tendency to prioritise cost-effective methods. Academia, on the other hand, shows a high percentage (93\%) and spreads across various attack methods. Our case study on the ByBit incident also exemplifies the complexity of real-world incidents compared to academic models. While academic literature often isolates attack vectors, the ByBit incident involved a multi-stage, multi-vector attack with a chain of sub-goals linked to the main goal of \teal{$sk$} compromise.

\subsection{High-Risk Third-Party Dependencies}

The ByBit attack highlights a critical systemic risk in modern wallet architectures: third-party dependencies can nullify even highly secure solutions. Despite ByBit’s use of hardware wallets, multi-sig authorisation, and transaction policies, its reliance on Safe’s third-party UI created a single point of failure. Similarly, Slope Wallet's reliance on a self-hosted instance of a third-party monitoring solution (Sentry) introduced vulnerabilities due to misconfiguration and operational errors. This further underscores how third-party integrations significantly impact wallet security. This demonstrates that wallet security inherits the weakest link in dependency chains. To mitigate these risks, wallets must adopt resilient architectures and proactively manage third-party risks through multi-layered audits and adversarial scenario modelling.

\subsection{Comparison of Custodial and Non-Custodial Attacks}
Our incident analysis reveals that custodial wallets and non-custodial accounts for 70\% and 30\% of attacks,  respectively.  Additionally, unknown methods are significantly higher in custodial wallets (50\%) than in non-custodial wallets (36\%). Incidents show a high degree of similarity between custodial and non-custodial attacks. For instance, in comparison to other attacks, phishing attacks account for a relatively high percentage of both custodial (10\%) and non-custodial (36\%) wallets, especially factoring in the number of unknown attacks. 

\subsection{High Malware and Phishing Attack Occurrence}

We also find that application attacks account for a significant percentage of incident occurrences (43\%), with 34\% in custodial wallets and 48\% in non-custodial wallets. Our data also indicates that malware and phishing attacks are the most common attack vectors, accounting for 10\% and 18\% of total incidents, respectively. We also find that phishing-malware attacks constitute 48\% of total non-custodial wallet attacks.

\subsection{Limitations of Security Measures}
\label{sec:def_dis_attacks}

The majority of defence implementations in academia are particularly tailored to specific advanced attacks such as \acs{puf} for microscopic attacks, correlation elimination sounds for non-invasive side channels, and \acs{puf} attacks. Despite this, academia does not account for sophisticated attacks, which may leverage multiple attack vectors. Furthermore, distributed architectures prevalent in the industry are insufficient if dependencies remain centralised. The ByBit breach demonstrates that security measures must extend to third-party components, requiring redundant safeguards such as on-chain transaction simulation to detect UI spoofing or logic hijacking. In addition, the Slope Wallet incident demonstrates how inadequate configuration of application monitoring tools can undermine otherwise secure implementations, highlighting the need for strict data scrubbing and monitoring configurations.

\subsection{Comparison of Precautionary and Remedial Defence Methods}
\label{sec:def_dis_attacks}

Our study presents defence methods applicable to various attack vectors, with the majority offering either precautionary or remedial strategies, as illustrated in \autoref{tab:defence_methods}. Notably, precautionary defences significantly outnumber remedial approaches, comprising roughly 89\% of all methods observed. Within the precautionary category, protection-focused implementations are the most prevalent, accounting for 58\%. Among remedial defences, detection methods are the most common at 17\%, while response and recovery measures each represent a mere 3\%. This disparity highlights a critical gap in reactive mitigation techniques, indicating a potential area for further development in response and recovery-focused defences.

\section{Discussion}
\label{sec:discussion}

\subsection{Limitations}

One significant limitation of our study is the quality and completeness of the data available on wallet attacks. As highlighted, many recorded incidents from custodial and non-custodial wallet providers contain a high degree of uncertainty regarding attack vectors (see \autoref{tab:attack-incidents}). This ambiguity restricts our capability to perform detailed quantitative analyses of wallet attacks, thereby limiting the precision of our analysis.

\subsection{Future Work}

To address these limitations, we propose the following research directions to improve wallet security:

\subsubsection{Enhanced Transaction Validation Measures} Our study highlights the uncertainty in recorded attack vectors, underscoring the need for enhanced transaction validation approaches. Advanced validation methods, such as independent transaction hash verification and proactive policy enforcement through on-chain gatekeeping, should be explored to improve transaction data clarity and reliability. Furthermore, integrating hardware wallets capable of clear-signing raw transaction parameters will significantly mitigate risks associated with deceptive UI interactions and unauthorised operational logic.

\subsubsection{Addressing Signature Verification Logic Flaws} Given the prevalence of signature verification logic flaws across wallet architectures, targeted research is crucial for developing secure and robust signature verification frameworks. Future work should prioritise the formal verification of signature verification algorithms, exploring cryptographic approaches specifically designed to mitigate known logic vulnerabilities. This will directly enhance the integrity and trustworthiness of wallet systems.

\subsubsection{Development of Reactive Defence Mechanisms} Our study identified a substantial gap in reactive security measures, with an evident imbalance favouring preventive strategies. Future research should emphasise the development of advanced reactive mitigation strategies, including real-time anomaly detection, responsive incident management protocols, and automated recovery frameworks tailored explicitly for wallet incidents. Enhancing reactive defence capabilities will substantially improve resilience and responsiveness to evolving threat vectors. By addressing these targeted research areas informed by our identified limitations, the community can significantly advance wallet security practices. This will lead to improved theoretical understanding and enhanced practical outcomes.

\section{Conclusion}
\label{sec:conclusion}

This paper systematically analyses the design, threats, attack vectors, and defensive strategies associated with cryptocurrency wallets. We introduce a comprehensive multi-dimensional taxonomy of wallet architectures, providing a structured and detailed framework to effectively understand and navigate the complex security landscape across various wallet types. By systematising diverse attack vectors, our framework offers clear insights into vulnerabilities and protective measures relevant to each wallet category.

Our analysis extends to examining 85 significant security incidents, accounting for financial losses exceeding \$6.98 billion. Through this systematic review, we propose targeted mitigation strategies corresponding to identified attack vectors and informed by our design taxonomy and security framework. Furthermore, our mapping of wallet mechanisms to specific design choices, threat profiles, attack methodologies, and existing defensive implementations underscores the critical interplay between different security dimensions and elucidates best practices.

We conduct a comparative analysis of incidents documented in industry contexts and vulnerabilities identified in academic research, revealing key gaps and convergence points between practical security threats and theoretical understandings. To further illustrate the practical applicability of our taxonomy and framework, we conduct detailed case studies, demonstrating its effectiveness in analysing and mitigating real-world wallet vulnerabilities and attacks.

By presenting an integrated perspective combining theoretical insights with empirical findings, our work lays the foundation for future research and practical advances, significantly enhancing the security and reliability of cryptocurrency wallets.















\section*{ACRONYMS}
\addcontentsline{toc}{section}{ACRONYMS}
\printacronyms[heading=none]

\section*{Acknowledgements}

The authors would like to thank Liyi Zhou and Zhipeng Wang for their in-depth review and constructive discussions during the study. This research is based upon work partially supported by the University Blockchain Research Initiative (UBRI)~\cite{feng2022university}. Any opinions, findings, and conclusions or recommendations expressed in this material are those of the authors and do not necessarily reflect the views of Ripple.

\bibliographystyle{cas-model2-names}
\bibliography{main}

\end{document}